\documentclass[%
superscriptaddress,
 amsmath,amssymb,
 aps,
floatfix,
prapplied, twocolumn
]{revtex4-2}

\usepackage{graphicx}
\graphicspath{{figs/}}
\usepackage{amsmath, amssymb}
\usepackage{mathtools}
\usepackage{multirow}

\newcommand{\eval}{\ensuremath \mathbf E}
\newcommand{\prob}{\ensuremath \mathbf P}
\newcommand{\probb}[1]{\prob\left(#1\right)}

\begin{document}

\title{Quantifying optical rogue waves}

\author{Éva Rácz}
\email{racz@optics.upol.cz}
\affiliation{Department of Optics, Palacký University, 17. listopadu 1192/12, 771 46 Olomouc, Czech Republic}
\author{Kirill Spasibko}
\affiliation{Q.ANT GmbH, Handwerkstraße 29, 70565 Stuttgart, Germany}
\author{Mathieu Manceau}
\affiliation{Laboratoire de Physique des Lasers, Université Sorbonne Paris Nord, CNRS, UMR 7538, F-93430 Villetaneuse, France}
\affiliation{Max Planck Institute for the Science of Light, Staudtstraße 2, 91058 Erlangen, Germany}
\author{László Ruppert}
\affiliation{Department of Optics, Palacký University, 17. listopadu 1192/12, 771 46 Olomouc, Czech Republic}
\author{Maria V.~Chekhova}
\affiliation{Max Planck Institute for the Science of Light, Staudtstraße 2, 91058 Erlangen, Germany}
\affiliation{Friedrich-Alexander Universität Erlangen-Nürnberg, Staudtstraße 7/B2, 91058 Erlangen, Germany}
\author{Radim Filip}
\affiliation{Department of Optics, Palacký University, 17. listopadu 1192/12, 771 46 Olomouc, Czech Republic}

\date{\today}

\begin{abstract}
This work presents two distinct approaches to estimating the exponent related to the distribution of optical rogue waves observed in supercontinuum generated in a single-mode fiber, that is, to quantifying the rogueness.
The first is a generalization of the well-known Hill estimator, and the second relies on estimating all parameters of a multi-parameter model. We show that the model shows a good correspondence with experimental data, and that the two estimating approaches provide consistent results, which are significantly more accurate than those obtained with earlier methods of estimation. 
Furthermore, alternative visualization through the tail function revealed the presence of pump depletion as well as detector saturation leading to the breakdown of power-law behavior for the largest observations. We characterized this breakdown via a combination of an exponential and a generalized Pareto distribution. Additionally, we have uncovered a weak memory effect in the data, which can be attributed to changes of the refractive index in the single-mode fiber. 
\end{abstract}

\maketitle

\section{Introduction}

Once only featured in sailors'---mostly disbelieved---stories, \emph{rogue waves} were first measured only relatively recently: in 1995 at an oil rig in the North Sea \cite{Sunde1995}. The concept, namely ocean waves that seem to appear ``out of nowhere'' and whose size far exceeds what is considered typical, has been observed or predicted in multiple fields where the wave analogy is applicable: Bose--Einstein condensates \cite{Bludov2009, Manikandan2014}, plasmas \cite{Ruderman2010, Moslem2011, Tsai2016}, atmospheric rogue waves \cite{Stenflo2010}, superfluids \cite{Ganshin2008}. Optical rogue waves were first reported in 2007 \cite{Solli2007}. The focus of theoretical research has been on providing generating mechanisms for such extreme behavior \cite{Buccoliero2011,Onorato2013,Hansen2021}. 

Instead of attempting to understand the various effects behind the emergence of rogue waves, the current work focuses on the statistical description of their magnitudes. The importance of this line of investigation lies in providing comparability: if such extreme behavior is to be harnessed, one should for quite practical reasons be able to answer the question of how large waves can be expected in a certain setup.  

Although with different points of emphasis, extreme and hard-to-predict behavior has been reported and studied considerably longer in other areas of research (see, for example, Pareto's seminal work on the distribution of wealth from 1896 \cite{Pareto1896}). Further examples include incomes \cite{Pareto1896,Yakovenko2009}, insurance claims \cite{Shpilberg1977,Rootzen1995}, number of citations of scientific publications \cite{Price1965,Redner1998,Golosovsky2017}, earthquake intensities \cite{Gutenberg1944,Christensen2002,Newberry2019}, avalanche sizes \cite{Birkeland2002}, solar flares \cite{Lu1991}, degree distributions of various social and biological networks \cite{Pastor2001,Stumpf2005,Newman2005}, and many more. Unsurprisingly, the mathematical background of extreme behavior has been most extensively studied with the motivation of mitigating financial risks (\hspace{1sp}\cite{Zipf1949, Mandelbrot1960, Mandelbrot1963, Embrechts1999, Rachev2003} and others). This broader area of research is referred to as extreme value theory, and \emph{heavy-tailed distributions} play an important role in it. These are defined as distributions whose tails decay slower than exponentially. 

A rogue wave is, by definition \cite{Akhmediev2010}, an unpredictably appearing wave with an amplitude at least twice as large than the significant wave height, with the tail of the amplitude probability density function (PDF) decaying slower that a Gaussian. Consequently, we are talking about extreme events with a heavier-than-Gaussian tail; that is, we can make use of the mathematical apparatus developed in extreme value theory. The aim of the current work is to estimate exactly how heavy the tail is, taking into account the particulars of the experimental setup. The simplest estimators for heavy tails \cite{Hill1975, Pickands1975, Kratz1996} build on the hypothesis that the distribution of interest is such that there exists a sharp, finite threshold beyond which the PDF decays exactly at a power-law rate. This is, of course, in most situations not true, and the basic estimators' bias can be reduced by taking into account higher-order behavior \cite{Feuerverger1999}. 

In comparison with more traditional uses of extreme value theory, non-linear optics \cite{Boyd2008, Grynberg2010} has clear advantages in controllability, reproducibility and statistical significance of the generated optical extreme events.  
Optical experiments producing light with heavy-tailed intensity distributions allow high repetition rates and therefore large samples to study unstable non-linear phenomena and their sources \cite{Barthelemy2008, Mercadier2009, Solli2008, Wetzel2012b}. However, obtaining the correct statistics of rogue-wave events is even so a non-trivial problem \cite{Sorensen2012,Wetzel2012}. Since the size of rogue events can exceed the median by many orders of magnitude \cite{Manceau2019}, their detection is difficult. Any physical measurement device has its limits; this is also the case for optical detectors: the detector response becomes non-linear and saturates with the increase of light power \cite{Quimby2006}. For example, for commonly-used biased photodiodes, the detector response cannot be larger than the reverse bias voltage; the further increase of input light power leads to the same output. Therefore, detector saturation will almost inevitably spoil the statistics of very energetic rogue events.

Furthermore, the efficiency of an optical non-linear process strongly depends on the intensity of its pump. This is also the case for four-wave mixing, which usually is the dominant effect producing supercontinuum generation. The amount of produced photons and the efficiency of the process depend exponentially on the pump intensity \cite{Agrawal2012}. Thus the efficiency of supercontinuum generation usually increases with the increase of pump power. However, as soon as the amount of converted pump energy becomes significant (tens of percent), the remaining pump cannot feed the process efficiently anymore; it is already depleted. In this case, the increase of the fiber length or the input pump power does not increase the total efficiency of the process \cite{Fedotov2003, Vanholsbeeck2005}.

The focus of the current work is on exploring two approaches to adapting the general statistics toolkit (proposed in \cite{Racz2021}) to the special case of rogue waves resulting from supercontinuum generation, using data collected during experiments similar to those reported in \cite{Manceau2019}. In this reference, an infinite-mean power-law distribution for supercontinuum generation was observed for the first time. However, a detailed statistical analysis of such phenomena requires complementary approaches. Furthermore, our investigation also uncovered some finer points of the behavior of supercontinuum generated from noisy light: we show the presence of pump depletion and a weak memory in the process.

\section{Methods}\label{sec:methods}

\subsection{Experiment}\label{sec:experiment}

\begin{figure}[hptb]
\centering
\includegraphics[width=8.6cm]{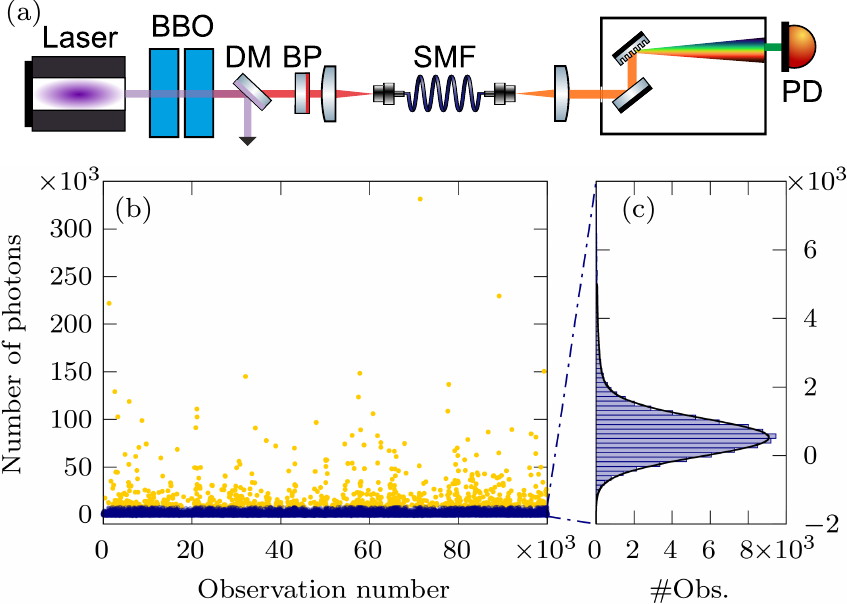}
\caption{Experimental setup and sample time series. (a) Schematics of the experimental setup. (b) Sample time series of the number of photons per pulse observed in an experiment. Time is shown in the number of the observation; the sampling frequency was 5 kHz. The bottom 99\% of photon number observations are colored blue, and the top 1\% are colored yellow. (c) Histogram based on the blue dots in (b). }\label{fig:experiment}
\end{figure}

In the experiment (Fig.\ \ref{fig:experiment}(a)), supercontinuum was obtained using a 5-m single-mode fiber (SMF) (Thorlabs P3-780A-FC-5) pumped by either bright squeezed vacuum (BSV) centered at 800 nm or by thermal light at 710 nm \cite{Manceau2019}. The energy per pulse was a few tens of nJ. Both BSV and thermal light were generated through type-I collinear parametric down-conversion (PDC), degenerate and non-degenerate, in two cascaded 3-mm barium borate (BBO) crystals from frequency-doubled radiation of a titanium-sapphire laser at 400 nm with a 1.4-ps pulse duration and up to $200\, {\mu}\mathrm{J}$ of energy per pulse.

After the laser radiation was cut off by a dichroic mirror (DM), either thermal light was injected into the fiber, by filtering radiation at \(710\pm 5\,\mathrm{nm}\), or BSV, by filtering it at \(800\pm 5\,\mathrm{nm}\). 
Subsequently, the supercontinuum was spectrally filtered by a monochromator with a 1-nm resolution, and measured by a photodetector (PD). The filtering was performed at 830 nm for the case of thermal light and at 760 nm for the case of BSV.
 The photodetector was calibrated such that its output voltage signal was converted into the number of photons per pulse, which is the data used in our analysis. The upper limit to the detector's linear response was \(\sim 10^6\) photons per pulse, with a maximum output \(\sim 2\cdot 10^6\) photons per pulse.

The lower part of Fig.~\ref{fig:experiment} shows a typical outcome of an experiment. In Fig.~\ref{fig:experiment}(b), we see the photon numbers measured by the detector as a function of time. The yellow points correspond to the top 1\% of observations, the blue points to the bottom 99\%. The histogram of the lower 99\% is shown in Fig.~\ref{fig:experiment}(c). From the histogram, one might mistakenly conclude that the interesting part of the distribution ends at about $5 \times 10^3$ photons per pulse. However, many observations in the top 1\% are significantly larger than the $99^{\mathrm{th}}$ percentile. This is the behavior that distinguishes these rogue waves from ``tamer'' processes, for which little of interest happens regarding the largest observations. For example, for an exponential sample of \(10^5\) observations, the expectancy of the sample maximum is only about 2.6 times larger than the \(99^{\mathrm{th}}\) percentile. In contrast, the same ratio for the particular measurement depicted in Fig.~\ref{fig:experiment} is about 41.

On the other hand, even though our detector has low noise and a decent dynamic range, the data at low photon numbers are affected by dark noise and at high numbers by detector saturation, which can be seen in the experimental probability distributions and should be considered during tail exponent estimation. 

\subsection{Terminology}\label{sec:terms}
The behavior of random variables can be described not only by the corresponding PDF or cumulative distribution function but also a lesser known alternative -- the tail function. 
For a real-valued random variable \(X\), this function can be defined as the probability that the value of the variable reaches or exceeds a pre-defined threshold \(x\):
\begin{equation}
\label{eq:tf}
\overline F(x) \coloneqq \probb{X \geqslant x},
\end{equation}
with \(\probb{\cdot}\) denoting the probability of an event. From a sample \(x_1, \ldots, x_n\), one can obtain the empirical tail function (ETF) simply as
\begin{equation}
\label{eq:etf-def}
\mathrm{ETF}(x) \coloneqq \frac{\#\left\{i: x_i \geqslant x \right\}}{n},
\end{equation}
with \(\#\{\cdot\}\) denoting the number of elements in the set.
We prefer using this function over the histogram because tail behavior is visually more apparent, and also because it does not depend on binning. 

The \emph{Pareto (or power-law) distribution} is the archetype of heavy-tailed distributions and can be given with the PDF \(f(x) = x_0^{-1}\left(x/x_0\right)^{-\alpha-1}\), for \(x_0, \alpha > 0\), and \(x \geq x_0\). The constant \(\alpha\) is referred to as the \emph{tail exponent} and as a measure of tail heaviness, it can be generalized to a broader class of distributions referred to as Pareto-type, or regularly varying distributions. Note that the statistical literature on heavy-tailed distributions predominantly uses the reciprocal of this exponent, \(\alpha^{-1}\) (known as the extreme value index), however, for the sake of simplicity, we will go on using the tail exponent. 

One of the problems regarding Pareto-type distributions is that some moments do not exist. For example, calculating empirical correlation becomes meaningless if the underlying distribution does not have a finite second moment. We would like to, however, detect correlations even for such rogue random variables. \emph{Rank correlation} \cite{Spearman1904,Hollander2013} is an alternative to regular correlation, and is defined even if no moments exist. It can be calculated as the usual linear correlation coefficient between the ranks of sets observations (i.e., only the ordering of the values matters, not the values itself). It is easy to show that for any random variable \(X\) with tail function \(\overline F(x)\), the random variable \(\overline F(X)\) is uniformly distributed on \([0,1]\) (so the mean is \(\frac 1 2\) and the variance is \(\frac 1 {12}\)). So with \(\overline G(y)\) standing for the tail function of a second random variable \(Y\), the definition of rank correlation can be given as 
\begin{equation}
\label{eq:rank-corr}
\varrho(X, Y) = 12\eval \left[\left(\overline F(X)-\frac 12\right) \left(\overline G(Y) - \frac 12\right)\right],
\end{equation}
with \(\eval(\cdot)\) denoting the expected value.\footnote{Note that the definition would be the same if one were to use the distribution function \(F(x) \equiv 1-\overline F(x)\) in the formula above.}

We can naturally define \emph{rank autocorrelation} in the same manner as regular autocorrelation, that is, as the rank correlation of the signal and a delayed copy: 
\begin{equation}\label{eq:rank-auto}
C(\delta) \coloneqq \varrho(X_i, X_{i+\delta}).
\end{equation}
Similarly to the usual definition of autocorrelation, calculating this function empirically produces only noise for independent and identically distributed data. Beside that this quantity is independent of the existence of moments, another advantage is that its value is not affected by strictly increasing transformations of the data, that is, in our case, detector saturation. This means that unlike for tail exponent estimation, it is unnecessary to discard the affected observations. 

\subsection{Tail exponent estimation}\label{sec:tailest}

In the supercontinuum intensity data to be analyzed, power-law behavior has both a lower and an upper limit. This upper limit is not considered in the statistical literature of tail exponent estimation because in the usual contexts, it is much harder to attain. In our case, it exists because the values of the largest observations are affected by detector saturation and pump depletion. In other words, even though the observed data is strictly speaking not heavy-tailed, we posit that this is due only to experimental limitations and would like to minimize their effect on exponent estimates. For this analysis, we present two separate approaches, which we will discuss in the following in detail.

\subsubsection*{Direct approach}

The first of these approaches we will refer to as the \emph{direct approach} because it estimates the value of the tail exponent directly, supposing that there is a finite interval within the range of observed values in which the density of the intensity observations shows a power-law decay. The advantages of this approach are that it is straightforward, and it is also more generally applicable than the current context of supercontinuum generation (it only assumes that there is an interval where the tail function decays at a power-law rate). However, it does not take into account deviations from exact power-law behavior (e.g., which might cause some bias for BSV source), nevertheless, it is an easy-to-implement tool that in our case has only little bias. 

The standard option for directly estimating the value of the tail exponent is the Hill estimator \cite{Hill1975}, defined as
\begin{equation}\label{eq:hill}
\hat \alpha^{-1}_{\mathrm H}(k) = \frac 1 k \sum_{i = 1}^{k}\ln x_{(i)} - \ln x_{(k+1)},
\end{equation}
with \(x_{(i)}\) denoting the \(i^{\mathrm{th}}\) largest element of the sample. Note that this formula is based on the \(k+1\) largest observations. However, since we have data that clearly do not decay at a power-law rate for large values, this method provides unreliable results (see Fig.\ \ref{fig:fits}, red line). Even if the tail of the distribution decays asymptotically at a power-law rate, for finite samples the problem of choosing \(k\) is non-trivial, and has an extensive literature \cite{Drees1998,Guillou2001,Danielsson2001,Caeiro2015}. The basic approach is visual and is referred to as the \emph{Hill plot}: one has to look for a range of values of \(k\) where \(\hat \alpha^{-1}_{\mathrm H}(k)\) is flat, that is, it is insensitive to the choice of \(k\). Note that choosing the tail length is equivalent to choosing a lower limit \(m\) beyond which observations are taken into account. That is, the tail length \(k\) can be expressed as \(k(m) = \max\{i: x_{(i)} > m\}\). We prefer plotting the parameter estimate as a function of the limit \(m\) because it makes for an easy comparison of different estimators (see Fig.~\ref{fig:sensitivity}(a)). 

In our previous work \cite{Racz2021}, we proposed a generalization of \eqref{eq:hill} for distributions for which power-law behavior has both a lower limit \(m\) and a finite upper limit \(M\). With \(k = \max\{i: x_{(i)} > m\}\) and \(j = \max\{i: x_{(i)} > M\}\), this generalized Hill estimator can be given as
\begin{align}
\nonumber \hat\alpha^{-1}_{\mathrm{gH}}(k, j)
 =& \frac{j}{k-j}\left(\ln x_{(j+1)}-\ln{x_{(k+1)}}\right) \\
 \label{eq:genhill} &+ \frac 1{k - j}\sum_{i = j+1}^k\left(\ln{x_{(i)}}-\ln{x_{(k+1)}}\right).
\end{align}

That is, out of the top \(k+1\) observations, one discards the \(j\) largest elements, and uses the remaining \(k+1-j\) observations to estimate the tail exponent; note that \(\hat \alpha^{-1}_{\mathrm H}(k) \equiv \hat\alpha^{-1}_{\mathrm{gH}}(k, 0)\). 

Clearly, with no prior information on \(m\) and \(M\), choosing their values based on the sample only is more involved than choosing the lower limit for the Hill estimator. We adapted an approach similar to the Hill plot, namely looking for an area where \(\hat\alpha_{\mathrm{gH}}\) is not sensitive to the choice of \([m, M]\). This can, for example, be done by plotting the value of \(\hat \alpha_{\mathrm{gH}}\) for several fixed values of \(M\) as a function of \(m\). The likely ranges for the limits can be pinpointed by looking at the ETF on a log-log scale. This visual approach is, of course, not feasible if one has a large number of samples to evaluate, but can be automated, for example, similarly to the heuristic algorithm proposed in \cite{Neves2015}.

\subsubsection*{Modeling approach}

The second approach, which we will refer to as the \emph{modeling approach}, consists of fitting a multi-parameter model to the whole process, where the tail exponent is only one parameter out of a few. The modeling approach discards only the largest observations, therefore it uses much more data points to obtain a fit, even if the majority of those data points are noisy. But if our noise model describes reality well, this approach can not only provide a better result, but in parallel, it explains and characterizes some of the most important experimental limitations. It is important to note that the proposed model is not explanatory in nature, but rather descriptive: its purpose is only to provide an alternative way to estimate the tail exponent. The experimental specifics and physical motivation of the model can be found in \cite{Manceau2019}.

We propose the following minimal model of the supercontinuum generation process:
\begin{equation}
\label{eq:model}
X_t = R\left[K\cdot\sinh^2\left(I_t + \omega_1 \right)\right]+\omega_2,
\end{equation}
with
\begin{itemize}
\item \(I_t\) standing for the incoming intensity with a constant mean \(\mu\), and
\begin{itemize}
\item exponentially distributed for thermal\\ (PDF \(\propto \exp\{-x/\mu\}\)),
\item gamma-distributed for a BSV source\\ (PDF \(\propto x^{-1/2}\exp\left\{-x/2\mu\right\}\));
\end{itemize}
\item \(K\): constant factor;
\item \(\omega_i \sim \mathcal N\left(0, \sigma_i^2\right)\): independent Gaussian noises; 
\item \(R(\cdot)\): detector response function.
\end{itemize}
The noise \(\omega_1\) corresponds to additive noises that affect the incoming intensity even before the light enters the fiber\footnote{Due to the \(\sinh^2(\cdot)\) transformation, this is essentially a multiplicative noise.}, whereas \(\omega_2\) is an additive detection noise.
In order to avoid introducing extra parameters for the response function, we did not fit the model to the observations affected by detector saturation. This amounted to only taking into account the non-linearity of detector response about the noise floor \(l\), through \(R(x) = \max\left\{l, x\right\}\), and discarding the largest observations.


This model can, of course, be further refined, but we were interested in the simplest version able to describe the observed process. This simplest version has five parameters: \(\vartheta = (\mu, l, K, \sigma_1, \sigma_2)\). The tail exponent of the output of this model is \(\alpha = (2\mu)^{-1}\) for the thermal, and \(\alpha = (4\mu)^{-1}\) for the BSV case. 


The advantage of the model defined by \eqref{eq:model} is that its density and distribution functions can be calculated semi-analytically. 
This gives us the opportunity to relatively easily perform a conditional (only observations below a pre-specified limit \(M\) are taken into account) maximum likelihood fit of its parameters. After the parameters are estimated, a simple binomial test can be performed to determine whether the fit should be rejected or not (further details of the method are discussed in \ref{sec:fit-details}).

\section{Results}\label{sec:results}
\begin{figure*}[t]
\centering
\includegraphics[width=0.95\textwidth]{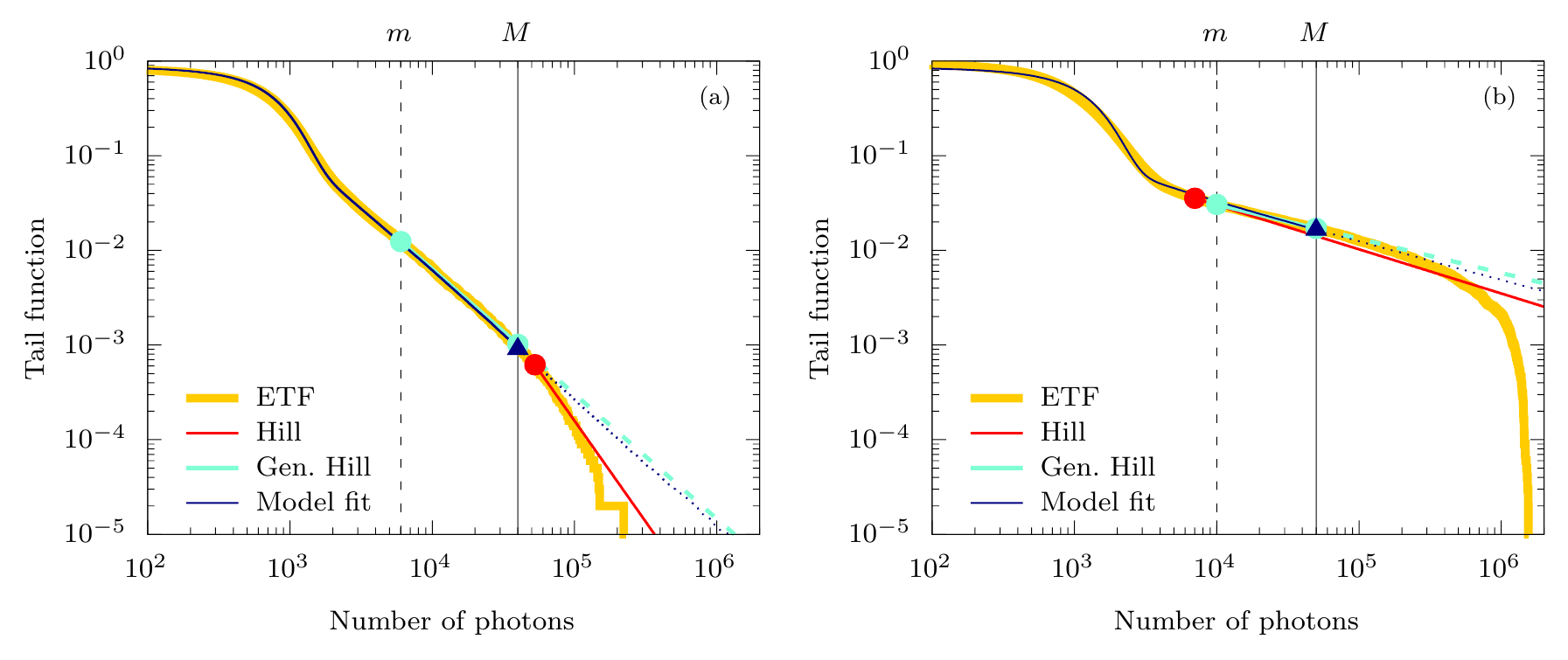}
\caption{Heavy tails in intensity fluctuations from a supercontinuum generation: comparison of different fitting approaches for (a) thermal, and (b) BSV pumping. The yellow line shows the empirical tail function (ETF) from an experimental data set. The red line shows the result of applying the traditional Hill estimator (\eqref{eq:hill}) combined with Guillou \& Hall's estimator \cite{Guillou2001} of the tail length \(k\), the red dot marks the beginning of the line. The light blue line (and its dashed extension) shows the result of the generalized version (\eqref{eq:genhill}) combined with a heuristic choice of the lower and upper cutoff parameters \(m\) and \(M\), the light blue dots mark the ends of the interval. The dark blue line (and its dotted extension) show the result of fitting the model \eqref{eq:model} to the sample; the value of the upper cutoff was the same as for the generalized Hill estimator and is marked by a dark blue triangle.}\label{fig:fits}
\end{figure*}

Our main aim in this paper was to estimate the exponent of the power-law decay in supercontinuum generation experiments. In the following, we will summarize the results we observed in supercontinuum generation pumped by thermal light and bright squeezed vacuum. We also compare our methods based on their obtained estimation and stability, and finally, show a memory effect generated during the process.


\subsection{Thermal source}\label{sec:res-thermal}

First, we investigate the properties of the tail function for intensities observed in supercontinuum generation experiments using a thermal source as a pump (Fig.~\ref{fig:fits}(a)). 
The theory \cite{Boyd2008, Manceau2019} 
suggests that in this case the tail should follow a power-law distribution, which, visually, means that the empirical tail function (at least for the largest values) should be linear on a log-log scale. However, experimental data (yellow) does not quite behave in that way: it seems linear only in the middle section. For low values (below $m$), this is due to the different types of noises, while for high values (above $M$), detector saturation and pump depletion affect the distribution in a significant way. It is important to note that even though we have quite large sample sizes (\(\sim 10^5\)), power-law behavior manifests itself only for the top few percent of observations, so the ``informative'' or effective sample size is considerably smaller (\(\sim  10^3\)).

The standard Hill estimator (\eqref{eq:hill}) in a realistic experiment is not the best choice of estimation method (red line) since the distribution clearly does not decay at a power-law rate for the largest values, which introduces a significant bias. A better direct estimation can be achieved after discarding the largest observations affected by saturation or pump depletion (that is, estimating the value of the tail exponent from the middle section only) by using the generalized Hill estimator from \eqref{eq:genhill} (light blue line). Finally, we have also used the modeling approach, consisting of constructing a detailed multi-parameter model of the process and estimating its parameters simultaneously, including the one that determines the value of the tail exponent (dark blue line). We can approximate the real process very accurately already with this simple model (the distribution for the lower values is not perfect, but as it can be seen in Fig.~\ref{fig:experiment}(c), it still provides a good fit, meaning that our model for noises is adequate for our purposes).

\begin{figure*}[hptb]
\centering
\includegraphics[width=0.95\textwidth]{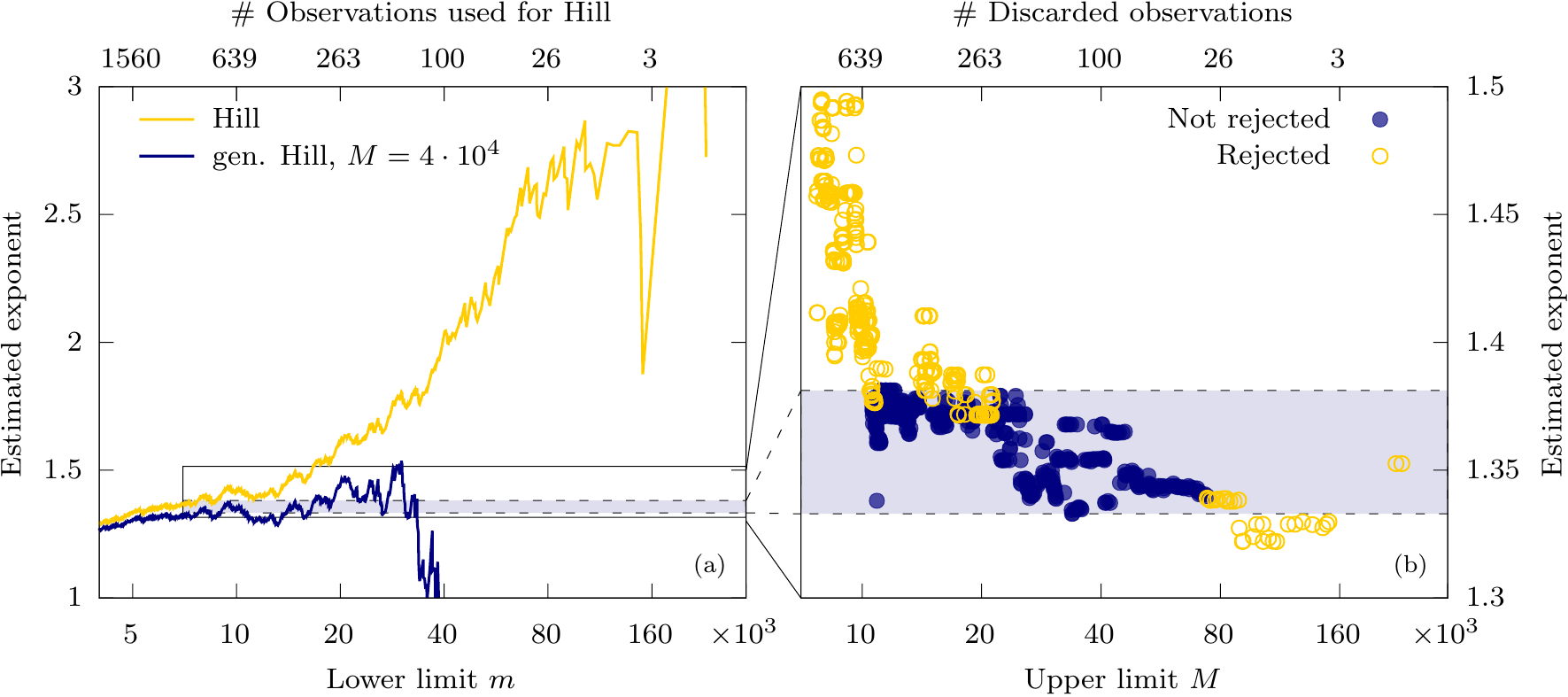}
\caption{Tail exponent estimates' dependence on cutoff choices. (a) Direct fitting methods: \(\hat \alpha_{\mathrm H}\left[k(m)\right]\) and \(\hat \alpha_{\mathrm gH}\left[k(m), j=100\right]\). 
(b) Modeling approach results as a function of the upper cutoff \(M\) (all observations below \(M\) were taken into account). The blue circles indicate the estimates that were accepted at a 5\% significance level by a binomial test, the estimates denoted by yellow circles were rejected (see \ref{sec:fit-details}).}
\label{fig:sensitivity}
\end{figure*}
\subsection{Bright squeezed vacuum}\label{sec:res-BSV}

If one uses bright squeezed vacuum (BSV) as a pump instead of thermal light to generate the supercontinuum, the situation remains quite similar (Fig.~\ref{fig:fits}(b)), but there are also a few significant differences. The difficulty comes from the fact that even though the asymptotic behavior is similar to the thermal case, the convergence to the asymptotics is slower. The Hill estimator (red line) is, as expected, biased as it includes the large, saturated values as well. But if we use the generalized Hill estimator (light blue line) with a properly chosen interval, we can obtain a reasonable estimate even in this case. Note however, that for this specific measurement, the range within which power-law decay is a reasonable assumption is quite short, about 1\%, of the data. The issue is that by using either version of the Hill estimator, one essentially estimates a gamma distribution with an exponential, which works well only asymptotically. However, the largest values are essentially lost due to saturation.

This is why the modeling approach can potentially provide more information. The fitted distribution (dark blue line) is not as close to the actual data as for a thermal source: The difference is quite visible for low values, which suggests that the model does not describe the experiment as precisely as for the thermal source. Nevertheless, since it is the power-law decay we are more interested in, accuracy about the noise floor is not crucial.
%



\subsection{Method comparison}\label{sec:res-analysis}

Let us highlight the sensitivities of the different methods to choosing the limit(s) of the application interval. Firstly, looking at the direct estimation methods (Fig.\ \ref{fig:sensitivity}(a)) using different lower limits, it becomes more evident why the value of the Hill estimator is rather incidental. For an ideal, strictly power-law distributed sample, the expectancy of the Hill estimator is constant, with its standard deviation decreasing as the number of points taken into account increases. For distributions that are only asymptotically power-law, there is an ideal tail length corresponding to a trade-off between minimizing bias and standard deviation. 

A simple approach to estimating this ideal value is looking for a plateau in the Hill plot \cite{Drees_HillPlot}, where the estimator's value is stable. Unfortunately, in our case, the Hill estimator (yellow line in Fig.~\ref{fig:sensitivity}(a)) presents a clear trend throughout the plotted interval as a function of \(m\) due to the bias introduced by the fact that the largest observations are clearly not Pareto-distributed. As a consequence, tail length estimators produce somewhat random values (red points in Fig.~\ref{fig:fits}). 


In contrast, throwing out the top observations really makes a difference compared to the standard Hill estimator: the generalized Hill estimator (blue line in Fig.~\ref{fig:sensitivity}(a)) is much less sensitive to the choice of the lower bound, that is, it produces relatively stable estimates in a wide range of values for \(m\). Note that compared to the traditional Hill estimator, the upper limit $M$ is an extra parameter to choose, which makes this method somewhat more complicated. But in practice, the estimator is also relatively stable in regard of choosing that parameter too, so it is not necessary to have a very accurate estimate of the endpoints of the interval \([m, M]\).

Looking at the sensitivity of the modeling approach (Fig.\ \ref{fig:sensitivity}(b)), we see that the estimates are even more stable with regard to the endpoint of the interval (in this case we only have an upper limit, \(M\)). More precisely, this approach provides a similar range of estimates as the generalized Hill estimator, but in addition, we can quantify the fit of the model with the estimated parameters to the measured data, and one can reject the values which are not a good fit (yellow points). And if we only look at the remaining, accepted estimates (blue points), these range only between 1.33 and 1.38 (light blue area), which is a really narrow interval, especially comparing it to the fluctuation of the traditional Hill estimates.

Of course, there is no guarantee that the given estimators are unbiased, so even if they are stable they could give a wrong result. That is why it is in general very useful to have at least two independent estimation methods (which are biased in different ways) and check whether they provide consistent results. In Table \ref{tab:fits} we can compare the results of different estimators, and can conclude that the modeling approach provides estimates close to the generalized Hill estimator. 

\begin{table}[h]
\centering
\begin{tabular}{c|c|c||c||c|c|c}
\multicolumn{3}{c||}{Thermal source} && \multicolumn{3}{c}{BSV source} \\ 
\hline
\multicolumn{2}{c|}{Direct} & \multirow{2}{*}{\begin{minipage}{1cm}\centering Our model\end{minipage}}& & \multicolumn{2}{c|}{Direct} & \multirow{2}{*}{\begin{minipage}{1cm}\centering Our model\end{minipage}}\\
\cline{1-2}\cline{5-6}
Hill & g. Hill & & & Hill & g. Hill &  \\
\hline\hline
2.13 & 1.32   & 1.37 & \(\hat \alpha\) & 0.47 & 0.36   & 0.37 \\
\hline
 0.23 & 0.38   & 0.37& \(\hat \mu\) & 0.54 & 0.69 & 0.68\\
- & -      & 490 &\(\hat l\)   & - & -       & 902\\
- & -      & 433& \(\hat K\)   & - & -     & 41\\
- & - &  0.54 & \(\hat \sigma_1\) & - & - & 0.83\\
 - & - & 526 & \(\hat \sigma_2\) & - & -  & 893
\end{tabular}
\caption{
Comparison of the results of different methods of fitting the value of the tail exponent \(\alpha\). The rows below the last horizontal line show the fitted values of the parameters of the model (\eqref{eq:model}). The notations \(\mu\), \(l\), \(K\), \(\sigma_1\), and \(\sigma_2\) stand for the mean pump intensity, the noise floor, the constant relating to the choice of the unit of output intensity, the (essentially) multiplicative noise, and the additive noise, respectively.}\label{tab:fits}
\end{table}

\subsection{Breakdown of power-law behavior}\label{sec:res-breakdown}

\begin{figure}[h]
\centering
\includegraphics[width=\columnwidth]{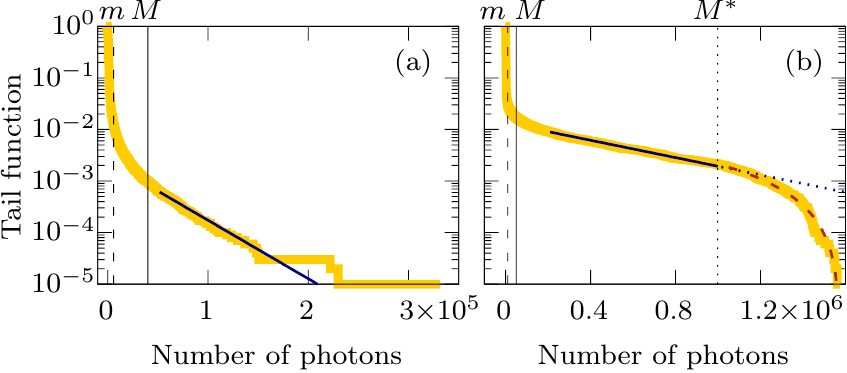}
\caption{Empirical tail functions on a semi-logarithmic scale: breakdown of power-law behavior. The samples are the same as shown in Fig.~\ref{fig:fits}, with (a) thermal, and (b) BSV pumping. The same values for the power-law interval \([m, M]\) are also indicated along the top horizontal axis. The dark blue solid lines show exponential fits, to observations above \(5\cdot 10^4\) in the thermal case, and between \(2\cdot 10^5\) and \(M^* = 10^6\) in the BSV case, providing exponential rate parameter estimates of \(\hat \lambda_a = 2.6  \cdot 10^{-5}\) and \(\hat \lambda_b = 1.9  \cdot 10^{-6}\), respectively. The dark blue dotted line in (b) is just the extension of the solid line beyond \(M^*\). The brown dashed line shows the generalized Pareto fit of the observations beyond \(M^*\), with \(\hat I_{\max} = 1.58\cdot 10^6\) and \(\hat \gamma = 1.59\).}\label{fig:large}
\end{figure}

As Fig.~\ref{fig:fits} shows, power-law behavior eventually breaks down; using a semi-logarithmic scale instead of the original log-log (Fig.~\ref{fig:large}) reveals exactly how: the linear sections in the plots indicate exponential decay rather than power-law (i.e., ETF \(\propto e^{-\lambda x}\) for some \(\lambda > 0\)). This behavior can be attributed to pump depletion: in the case of a fluctuating pump, the depletion manifests itself not only in the power reduction and decrease of generation efficiency but also in the change of pump statistics \cite{Manceau2019, Florez2020}. Since the process efficiency is higher for more energetic pump bursts, more energy is converted, particularly from these bursts, reducing pump fluctuations and changing their statistics. Furthermore, since all these processes happen simultaneously during the propagation in the optical fiber, deriving a rigorous theoretical framework is a complex task. Our model is therefore limited to the undepleted pump approximation, constant pump power, and constant pump statistics along the propagation in the fiber.

Furthermore, as Fig.~\ref{fig:large}(b) shows, if intensities are high enough (beyond about \(M^*=10^6\)), the decay becomes even faster than exponential due to detector saturation. It can be shown (see \ref{sec:beyond}) that combining an exponential distribution with an exponentially saturating detector results in a generalized Pareto distribution. More accurately, the ETF is \(\propto \left(I_{\max} - x\right)^\gamma\). Here, \(I_{\max}\) denotes the maximum output of the detector, and \(\gamma > 0\) the related exponent. 

The parameters of the exponential (\(\lambda\)) and the generalized Pareto distributions (\(I_{\max}\) and \(\gamma\)) can be estimated in a similar fashion to the tail exponent via a conditional maximum likelihood approach (see \ref{sec:beyond}). The results of the exponential fits are shown in blue in Fig.~\ref{fig:large}, the result of the generalized Pareto fit is shown in Fig.~\ref{fig:large}(b) using a brown dashed line.

As an interesting note, we mention that the generalized Pareto distribution (ETF \(\propto \left(1+\xi x \right)^{-1/\xi})\) includes the regular Pareto distribution (\(\xi > 0\)), the exponential distribution (\(\xi = 0\)), and the distribution derived for the observations beyond \(M^*\) (\(\xi < 0\)) as well, so it should be possible to introduce a decreasing \(\xi(x)\) function to treat the three types of behavior at the same time. 
Moreover, in principle, one could include these aspects in the modeling approach of estimating the tail exponent as well, however, as the tail exponent has only little effect on the few observations beyond \(M\), doing so would not likely improve the tail exponent estimation. On the contrary, introducing the extra parameters would negatively affect the stability of the numerical optimization.

\subsection{Memory effect}\label{sec:res-memory}



The discussed methods rely on the assumption of independent and identically distributed observations. This assumption should be checked empirically, even if we think it is true according to our understanding of the physical process. As discussed in Sec.~\ref{sec:methods}.\ref{sec:terms}, quantifying interdependence is non-trivial if the underlying random variables do not have a finite second moment; thus, rank correlations \eqref{eq:rank-auto} are used instead of regular correlations.  

\begin{figure}[hptb]
\centering
\includegraphics[width=\columnwidth]{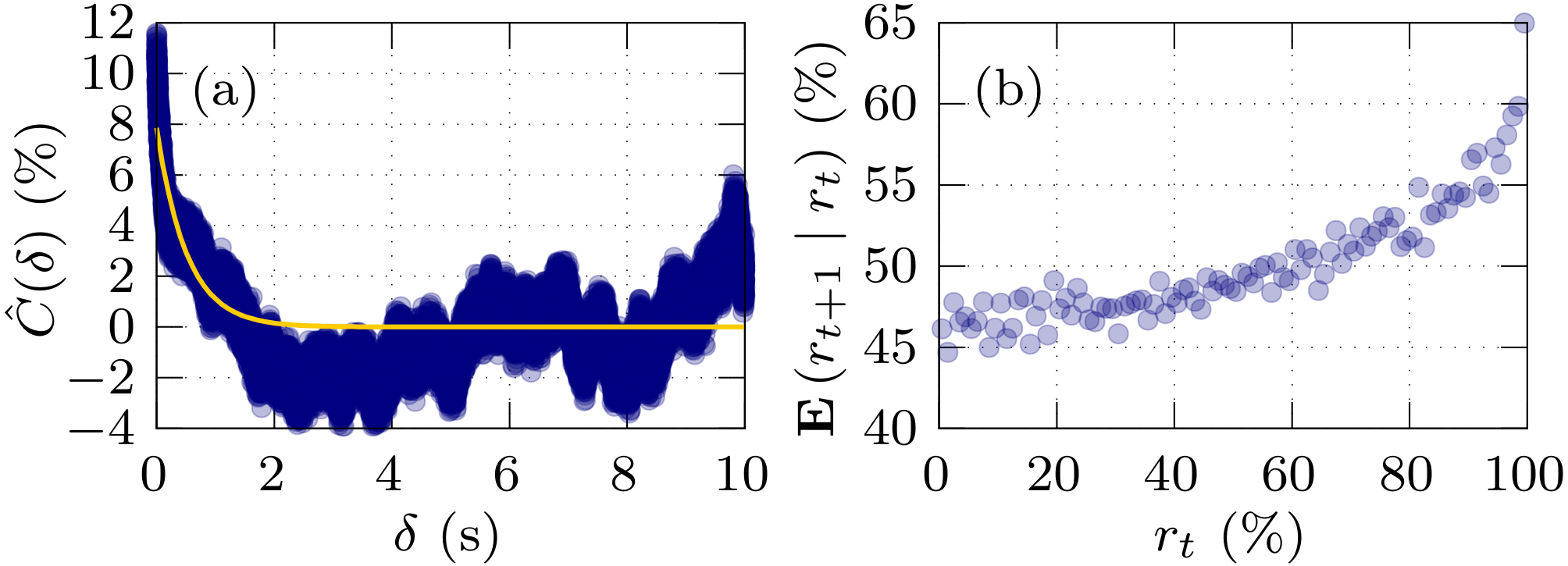}
\caption{Memory effect in supercontinuum generation. (a) Rank autocorrelation's dependence on the time-lag \(\delta\) between observations. The yellow line shows the fitted exponential decay. (b) Average rank of an observation conditional on the percentile of the previous observation.}\label{fig:memory}
\end{figure}

Figure~\ref{fig:memory}(a) shows the empirical rank autocorrelation for a BSV-pumped sample. This particular sample was chosen because it displays stronger than average correlations, but the general behavior is similar in most of our samples: positive rank correlation between subsequent observations on the order of a few percents, which decays with a correlation time of \(\sim 1\,\mathrm{s}\). There are also some samples that do not display any significant correlation, and some that display periodicity with a period of a few seconds.

As a further measure of correlations, in Fig.~\ref{fig:memory}(b), we show the average rank of events conditional on the rank of the previous event (same sample as in Fig.~\ref{fig:memory}(a)). We chose this particular plot because it does not only show that a correlation is present, but it also reveals that high-photon-number pulses are more likely to be bunched together than smaller ones (note that since we are using ranks instead of observed values the distortion of the detector is automatically eliminated). 
We assume that the cause of this bunching is connected to the photorefractive effect in the fiber \cite{Hand1990}.  
In any case, it is clear that the observed samples are not quite memoryless. However, the calculated correlations are not very high (even the correlation between subsequent observations is typically lower than 5\%), which is why either of our suggested approaches can be deemed applicable.

\section{Conclusion and Discussion}\label{sec:conclusion}

Optical rogue waves present an exciting direction of research, and provide excellent means to study extreme statistical behavior in a controllable and reproducible fashion on large amounts of data. This study is relevant to nonlinear optics and light-matter interaction, and also simulates extreme events which are statistically rare but still present in other fields of science. We analyzed intensity data gathered from supercontinuum generation experiments fed by thermal light and bright squeezed vacuum. 

The rogueness of the observed intensity distributions comes from the fact that they display a power-law decay. We used two different, advanced methods to estimate the related exponent. Accurately estimating the rate of the power-law decay is of both theoretical and practical importance. Knowing whether the theoretical distribution (unaffected by saturation issues) has a finite second moment or not determines whether, for example, calculating the correlation function \(g^{(2)}\) is meaningful, and having a proper estimate of the tail exponent also helps with designing further experiments.

The first method estimates the value of the tail exponent directly, and is the generalization of the well-known Hill estimator. As pump depletion and detector saturation have an effect on the largest values only, the estimation of the tail is much more precise if we discard these affected observations. This requires a modification of the Hill estimator, but is otherwise quite straightforward.
The second approach consists of devising a simplified physical model of the process (including noises and further limitations), and performing a maximum likelihood fit of its parameters based on the data. This approach is, of course, tailored to the specific problem at hand, and in a different context, requires setting up a different model. The largest values are discarded similarly to the first approach, which means that instead of having to add extra parameters to the model to describe pump depletion and the saturation curve of the detector, we only need to choose a limit beyond which values are discarded. Since the fit is quite good, this means we not only accurately estimated the parameters (including the tail exponent), but we also have a reasonably good grasp of the imperfections in the system. Using the two approaches in parallel is useful as well, since they produce independent estimates. Consistent values indicate that our view of the process, and our implementation of the estimators are appropriate. 

This dual approach is especially important in the case of pumping with bright squeezed vacuum. We have a fundamental problem in both our estimation schemes: the direct method is better suited for an underlying exponential distribution than for gamma, while the modeling approach has problems with the fit as the physical process is more complex. Nevertheless, either method provides similar tail exponents, meaning that albeit the estimation procedures are more sensitive in the BSV case, we can still perform a consistent tail estimation. 

Other than estimating the exponent of the power-law decay, we also showed how power-law behavior breaks down for the largest observations. The primary cause of this is pump depletion, turning the power-law decay into exponential. Furthermore, especially in the case of pumping with bright squeezed vacuum, high intensities beyond the detector's linear response range were achieved, which additionally distorted the empirical distribution. We were able to provide a model to characterize this post-power-law range of observations through an exponential and a generalized Pareto distribution.

Furthermore, the process displays a moderate memory effect. We assume this is connected to the photorefractive effect in the optical fiber, and is stronger for larger observations. It causes the bunching of high-energy pulses in time, which can be more or less prominent depending on the settings. Overall, it is certainly an interesting phenomenon that requires a more detailed investigation.

\begin{acknowledgments}
É.~R.\ and L.~R.\ acknowledge the support of project 19-22950Y of the Czech Science Foundation. R.~F.\ acknowledges the support of project 21-13265X of the Czech Science Foundation. As members of project SPARQL,  L.~R. and É.~R. acknowledge funding from the MEYS of the Czech Republic (Grant Agreement 8C22001), M.~V.~C. acknowledges funding from the Deutsche Forschungsgemeinschaft (grant number CH 1591/16-1). Project SPARQL has received funding from the European Union's Horizon 2020 Research and Innovation Programme under Grant Agreement no. 731473 and 101017733 (QuantERA). L.~R., É.~R. and R.~F. have further been supported by the European Union’s 2020 research and innovation programme (CSA - Coordination and support action, H2020-WIDESPREAD-2020-5) under grant agreement No.~951737 (NONGAUSS). 
\end{acknowledgments}

\appendix
\section{Modeling approach details}\label{sec:fit-details}
The distribution of our model \eqref{eq:model} can be calculated semi-analytically. More specifically, the PDF \(f_1\) corresponding to adding a Gaussian noise \(\omega_1\) to the exponentially/gamma distributed pump intensity can be given in closed form. Consequently, applying the \(\max\left\{K\cdot \sinh^2(\cdot), l\right\}\) transformation to it, with which we model the effect of the four-wave mixing process in the single-mode fiber, can also be given in a closed form, which we denote by \(f_2\). It is only the convolution of \(f_2\) with the second Gaussian that has to be performed numerically to obtain the PDF \(f_3(\cdot)\) corresponding to the full model. 

The first PDF can be given as 
\begin{align}
\nonumber f_{1}^{\text{th.}}(\mu,\sigma_1; x) &=
 \sigma_1^{-1}\varphi\left(\frac x {\sigma_1}\right) + \mu^{-1} e^{-\frac x \mu -\frac{\sigma_1^2}{2\mu}}\cdot \Phi\left(\frac x {\sigma_1}- \frac {\sigma_1^2}{\mu}\right) \\
\label{eq:f1-th} &- \sigma_1^{-1}e^{-\frac x \mu -\frac{\sigma_1^2}{2\mu}}\cdot \varphi\left(\frac x {\sigma_1} - \frac {\sigma_1^2}{\mu}\right),
\end{align}
for the thermal case, with \(\varphi(\cdot)\) and \(\Phi(\cdot)\) denoting the standard normal density and cumulative distribution function, respectively. For the BSV pumping, we have
\begin{align}
\nonumber f_{1}^{\text{BSV}}(\mu, \sigma_1; x) =&
e^{-\frac{x}{2\mu}-\frac {a^2} 4} 
\sqrt{\frac{|a|}{16\mu \sigma_1}} \\
\label{eq:f1-BSV-a}&\times
\left\{
\begin{array}{ll}
I_{\frac{1}{4}}\left(\frac{a^2}{4}\right)+I_{-\frac{1}{4}}\left(\frac{a^2}{4}\right), & x > \frac{\sigma_1^2}{2\mu}\\
\frac {\sqrt 2}{\pi} K_{\frac{1}{4}}\left(\frac{a^2}{4}\right), & x < \frac{\sigma_1^2}{2\mu}
\end{array}
\right.  \\
\label{eq:f1-BSV-b}
f_{1}^{\text{BSV}}\left(\mu, \sigma_1;\frac{\sigma_1^2}{2\mu}\right) =& \frac{\sqrt[4]{2} \Gamma \left(\frac{5}{4}\right) e^{-\left(\frac{\sigma_1}{2 \mu }\right)^2}}{\pi  \sqrt{\mu  \sigma_1 }}, 
\end{align}
with \(a \equiv \frac {x}{\sigma_1} - \frac{\sigma_1}{2\mu}\), and \(I_{\nu}(\cdot)\), \(K_{\nu}(\cdot)\) denoting the modified Bessel functions of the first and second kind, respectively. 

The second PDF can be given for \(x \geqslant l\) as
\begin{align}
\nonumber f_2(\mu, l, K, \sigma_1; x) &= 
\frac {f_1\left(\sinh^{-1}\sqrt{\frac x K}\right)+f_1\left(-\sinh^{-1}\sqrt{\frac x K}\right)}
{2 \sqrt{x(x+K)}} \\
\label{eq:f2} &+ 
\delta(l)\int_{-\sinh^{-1}\sqrt{\frac l K}}^{\sinh^{-1}\sqrt{\frac l K}}f_1(u)\, \mathrm d u
\end{align}
with \(\delta(\cdot)\) denoting the Dirac delta function; for \(x< l\), \(f_2(\mu, l, K, \sigma_1;x) = 0\).

The PDF \(f_2\) can be convoluted numerically with a Gaussian distribution with variance \(\sigma_2\) to obtain the PDF of the model \eqref{eq:model}. We denote this resulting PDF with \(f_3(\vartheta; x)\), using a shorthand for the set of parameters: \(\vartheta \equiv \left\{\mu, l, K, \sigma_1, \sigma_2\right\}\). As we are well aware that for the largest values, the model does not hold due to pump depletion and detector saturation, we use a conditional maximum likelihood estimator. The maximum likelihood estimator is conditional in that only observations below a pre-specified limit \(M\) (beyond which the behavior is markedly not power-law) are taken into account, that is, the objective function is
\begin{equation}
\label{eq:objective}
Obj\left(M; \vartheta\right) = 
\prod_{i :  x_i < M} \frac{f_3(\vartheta; x_i)}{F_3(\vartheta; M)},
\end{equation}
with \(F_3(\vartheta; \cdot)\) denoting the cumulative distribution corresponding to \(f_3(\vartheta; \cdot)\). That is, given a value \(M\) for the upper limit, and restricting ourselves to the observations \(x_i\) below it, we look for the set of parameters \(\hat\vartheta\) that maximizes \eqref{eq:objective}.
Note that objective functions for different values of \(M\) are not comparable as they rely on different numbers of observations.

Unlike direct estimation, this modeling approach also says something regarding the discarded observations: namely how many observations there should be beyond \(M\). This allows for a simple binomial test: for a set of parameters \(\hat\vartheta\) that maximized the objective function in \eqref{eq:objective} for a given value of \(M\), one can calculate \(p^* = 1- F_3(\hat\vartheta; M)\), which is the probability that observations reach or exceed \(M\). Given a total sample size of \(n\), the number of observations \(n_M\) beyond \(M\) is binomially distributed, \(n_M \sim \mathrm{BIN}(n, p^*)\). Therefore, given the fit represented here only by \(p^*\), it is easy to assess how likely it is to have as many observations beyond \(M\) as the data presents. 

\section{Beyond the power-law}\label{sec:beyond}
\setcounter{equation}{0}
As Fig.~\ref{fig:large} shows, there is an exponentially decaying interval of observations (ETF \(\propto e^{-\lambda x}\)) in both the thermal and the BSV samples, corresponding to pump depletion. Figure~\ref{fig:large}(b) displays the effect of detector saturation on top of pump depletion. The aim of this section is to show how applying detector saturation to an exponentially decaying distribution changes the statistics, that is, we provide a parametric model for the observations beyond \(M^*\) in Fig.~\ref{fig:large}(b). We also briefly describe how to estimate the parameters corresponding to the exponentially decaying interval and the saturating interval as well. 

According to direct measurements of the detector's response to increasing intensities, the behavior close to the maximum response $I_{\max}$ is described well by an exponential saturation: \(I_{\mathrm{out}} = R(I_{\mathrm{in}}) = I_{\max} - C\cdot \exp\left\{-I_{\mathrm{in}}/\Delta\right\}\),
for \(I_{\max}  > I > I_1\). 
Substituting this response function into the ETF of the detected intensity yields \(\mathbf P\left(I_{\text{out}} \geqslant x\right) \propto \mathbf P\left(I_{\text{in}} \geqslant R^{-1}(x)\right) \propto \left(I_{\max} - x\right)^\gamma\), with \(\gamma = \lambda \Delta\). This is a generalized Pareto distribution (GPD) \cite{Pickands1975} with shape parameter \(\xi = -1/\gamma\).  

\paragraph{Exponential fit}
The formulas for the maximum likelihood fits for the exponential portion of the distribution are very similar to direct power-law fits, as the logarithm of a power-law distributed random variable is exponentially distributed. Therefore, one only has to remove the logarithms from \eqref{eq:genhill} to obtain
\begin{align}
\nonumber \hat\lambda^{-1}(k, j)
 =& \frac{j}{k-j}\left(x_{(j+1)}-{x_{(k+1)}}\right) \\
 \label{eq:estexp} &+ \frac 1{k - j}\sum_{i = j+1}^k\left({x_{(i)}}-{x_{(k+1)}}\right)
\end{align}
as an estimator of the rate parameter \(\lambda\) of an exponential distribution with PDF \(\propto e^{-\lambda x}\). 

\paragraph{Generalized Pareto fit}
The parameters of the GPD can be estimated via conditional maximum likelihood estimation based on the \(k+1\) largest observations. The estimate \(\hat I_{\max}(k)\) is the solution of the equation
\begin{equation}
\label{eq:est-imax}
1 = \left(1 - \frac 1 k \sum_{j = 1}^k \ln \frac{I_{\max} - x_{(k+1)}}{I_{\max} - x_{(j)}}\right)
\cdot \frac 1 k \sum_{j = 1}^k \frac{I_{\max} - x_{(k+1)}}{I_{\max} - x_{(j)}},
\end{equation}
with \(x_{(j)}\) denoting the \(j^{\mathrm{th}}\) largest observation in the sample. Having numerically solved the previous equation for \(I_{\max}\), the estimate \(\hat \gamma(k)\) can be calculated as
\begin{equation}
\label{eq:est-gamma}
\hat\gamma^{-1}(k) = \frac 1 k \sum_{j = 1}^k \ln \frac{\hat I_{\max}(k) - x_{(k+1)}}{\hat I_{\max}(k) - x_{(j)}}.
\end{equation}
Note that similarly to the Hill estimator, this estimate also depends on the choice of the tail length \(k\). 
One can choose it in the manner of evaluating the Hill plot, that is, by looking for a plateau in the estimates. The result of the GPD fit is shown in Fig.~\ref{fig:large}(b) with a dashed brown line, the exponential fits are blue.

\bibliography{heavy_tailed.bib}

\begin{thebibliography}{63}%
\makeatletter
\providecommand \@ifxundefined [1]{%
 \@ifx{#1\undefined}
}%
\providecommand \@ifnum [1]{%
 \ifnum #1\expandafter \@firstoftwo
 \else \expandafter \@secondoftwo
 \fi
}%
\providecommand \@ifx [1]{%
 \ifx #1\expandafter \@firstoftwo
 \else \expandafter \@secondoftwo
 \fi
}%
\providecommand \natexlab [1]{#1}%
\providecommand \enquote  [1]{``#1''}%
\providecommand \bibnamefont  [1]{#1}%
\providecommand \bibfnamefont [1]{#1}%
\providecommand \citenamefont [1]{#1}%
\providecommand \href@noop [0]{\@secondoftwo}%
\providecommand \href [0]{\begingroup \@sanitize@url \@href}%
\providecommand \@href[1]{\@@startlink{#1}\@@href}%
\providecommand \@@href[1]{\endgroup#1\@@endlink}%
\providecommand \@sanitize@url [0]{\catcode `\\12\catcode `\$12\catcode
  `\&12\catcode `\#12\catcode `\^12\catcode `\_12\catcode `\%12\relax}%
\providecommand \@@startlink[1]{}%
\providecommand \@@endlink[0]{}%
\providecommand \url  [0]{\begingroup\@sanitize@url \@url }%
\providecommand \@url [1]{\endgroup\@href {#1}{\urlprefix }}%
\providecommand \urlprefix  [0]{URL }%
\providecommand \Eprint [0]{\href }%
\providecommand \doibase [0]{https://doi.org/}%
\providecommand \selectlanguage [0]{\@gobble}%
\providecommand \bibinfo  [0]{\@secondoftwo}%
\providecommand \bibfield  [0]{\@secondoftwo}%
\providecommand \translation [1]{[#1]}%
\providecommand \BibitemOpen [0]{}%
\providecommand \bibitemStop [0]{}%
\providecommand \bibitemNoStop [0]{.\EOS\space}%
\providecommand \EOS [0]{\spacefactor3000\relax}%
\providecommand \BibitemShut  [1]{\csname bibitem#1\endcsname}%
\let\auto@bib@innerbib\@empty
\bibitem [{\citenamefont {Sunde}(1995)}]{Sunde1995}%
  \BibitemOpen
  \bibfield  {author} {\bibinfo {author} {\bibfnamefont {A.}~\bibnamefont
  {Sunde}},\ }\bibfield  {title} {\bibinfo {title} {Kjempeb\o{}lger i
  {N}ordsj\o{}en ({E}xtreme waves in the {N}orth {S}ea)},\ }\href@noop {}
  {\bibfield  {journal} {\bibinfo  {journal} {V{\ae}r {\&} Klima}\ } (\bibinfo
  {year} {1995})},\ \bibinfo {note} {(in Norwegian)}\BibitemShut {NoStop}%
\bibitem [{\citenamefont {Bludov}\ \emph {et~al.}(2009)\citenamefont {Bludov},
  \citenamefont {Konotop},\ and\ \citenamefont {Akhmediev}}]{Bludov2009}%
  \BibitemOpen
  \bibfield  {author} {\bibinfo {author} {\bibfnamefont {Y.~V.}\ \bibnamefont
  {Bludov}}, \bibinfo {author} {\bibfnamefont {V.~V.}\ \bibnamefont
  {Konotop}},\ and\ \bibinfo {author} {\bibfnamefont {N.}~\bibnamefont
  {Akhmediev}},\ }\bibfield  {title} {\bibinfo {title} {Matter rogue waves},\
  }\href {https://doi.org/10.1103/PhysRevA.80.033610} {\bibfield  {journal}
  {\bibinfo  {journal} {Phys. Rev. A}\ }\textbf {\bibinfo {volume} {80}},\
  \bibinfo {pages} {033610} (\bibinfo {year} {2009})}\BibitemShut {NoStop}%
\bibitem [{\citenamefont {Manikandan}\ \emph {et~al.}(2014)\citenamefont
  {Manikandan}, \citenamefont {Muruganandam}, \citenamefont {Senthilvelan},\
  and\ \citenamefont {Lakshmanan}}]{Manikandan2014}%
  \BibitemOpen
  \bibfield  {author} {\bibinfo {author} {\bibfnamefont {K.}~\bibnamefont
  {Manikandan}}, \bibinfo {author} {\bibfnamefont {P.}~\bibnamefont
  {Muruganandam}}, \bibinfo {author} {\bibfnamefont {M.}~\bibnamefont
  {Senthilvelan}},\ and\ \bibinfo {author} {\bibfnamefont {M.}~\bibnamefont
  {Lakshmanan}},\ }\bibfield  {title} {\bibinfo {title} {Manipulating matter
  rogue waves and breathers in {B}ose-{E}instein condensates},\ }\href
  {https://doi.org/10.1103/PhysRevE.90.062905} {\bibfield  {journal} {\bibinfo
  {journal} {Phys. Rev. E}\ }\textbf {\bibinfo {volume} {90}},\ \bibinfo
  {pages} {062905} (\bibinfo {year} {2014})}\BibitemShut {NoStop}%
\bibitem [{\citenamefont {Ruderman}(2010)}]{Ruderman2010}%
  \BibitemOpen
  \bibfield  {author} {\bibinfo {author} {\bibfnamefont {M.~S.}\ \bibnamefont
  {Ruderman}},\ }\bibfield  {title} {\bibinfo {title} {Freak waves in
  laboratory and space plasmas},\ }\href
  {https://doi.org/10.1140/epjst/e2010-01238-7} {\bibfield  {journal} {\bibinfo
   {journal} {The European Physical Journal Special Topics}\ }\textbf {\bibinfo
  {volume} {185}},\ \bibinfo {pages} {57} (\bibinfo {year} {2010})}\BibitemShut
  {NoStop}%
\bibitem [{\citenamefont {Moslem}\ \emph {et~al.}(2011)\citenamefont {Moslem},
  \citenamefont {Shukla},\ and\ \citenamefont {Eliasson}}]{Moslem2011}%
  \BibitemOpen
  \bibfield  {author} {\bibinfo {author} {\bibfnamefont {W.~M.}\ \bibnamefont
  {Moslem}}, \bibinfo {author} {\bibfnamefont {P.~K.}\ \bibnamefont {Shukla}},\
  and\ \bibinfo {author} {\bibfnamefont {B.}~\bibnamefont {Eliasson}},\
  }\bibfield  {title} {\bibinfo {title} {Surface plasma rogue waves},\ }\href
  {https://doi.org/10.1209/0295-5075/96/25002} {\bibfield  {journal} {\bibinfo
  {journal} {Europhysics Letters}\ }\textbf {\bibinfo {volume} {96}},\ \bibinfo
  {pages} {25002} (\bibinfo {year} {2011})}\BibitemShut {NoStop}%
\bibitem [{\citenamefont {Tsai}\ \emph {et~al.}(2016)\citenamefont {Tsai},
  \citenamefont {Tsai},\ and\ \citenamefont {I}}]{Tsai2016}%
  \BibitemOpen
  \bibfield  {author} {\bibinfo {author} {\bibfnamefont {Y.-Y.}\ \bibnamefont
  {Tsai}}, \bibinfo {author} {\bibfnamefont {J.-Y.}\ \bibnamefont {Tsai}},\
  and\ \bibinfo {author} {\bibfnamefont {L.}~\bibnamefont {I}},\ }\bibfield
  {title} {\bibinfo {title} {Generation of acoustic rogue waves in dusty
  plasmas through three-dimensional particle focusing by distorted waveforms},\
  }\href {https://doi.org/10.1038/nphys3669} {\bibfield  {journal} {\bibinfo
  {journal} {Nature Physics}\ }\textbf {\bibinfo {volume} {12}},\ \bibinfo
  {pages} {573} (\bibinfo {year} {2016})}\BibitemShut {NoStop}%
\bibitem [{\citenamefont {Stenflo}\ and\ \citenamefont
  {Marklund}(2010)}]{Stenflo2010}%
  \BibitemOpen
  \bibfield  {author} {\bibinfo {author} {\bibfnamefont {L.}~\bibnamefont
  {Stenflo}}\ and\ \bibinfo {author} {\bibfnamefont {M.}~\bibnamefont
  {Marklund}},\ }\bibfield  {title} {\bibinfo {title} {Rogue waves in the
  atmosphere},\ }\href {https://doi.org/10.1017/S0022377809990481} {\bibfield
  {journal} {\bibinfo  {journal} {Journal of Plasma Physics}\ }\textbf
  {\bibinfo {volume} {76}},\ \bibinfo {pages} {293–295} (\bibinfo {year}
  {2010})}\BibitemShut {NoStop}%
\bibitem [{\citenamefont {Ganshin}\ \emph {et~al.}(2008)\citenamefont
  {Ganshin}, \citenamefont {Efimov}, \citenamefont {Kolmakov}, \citenamefont
  {Mezhov-Deglin},\ and\ \citenamefont {McClintock}}]{Ganshin2008}%
  \BibitemOpen
  \bibfield  {author} {\bibinfo {author} {\bibfnamefont {A.~N.}\ \bibnamefont
  {Ganshin}}, \bibinfo {author} {\bibfnamefont {V.~B.}\ \bibnamefont {Efimov}},
  \bibinfo {author} {\bibfnamefont {G.~V.}\ \bibnamefont {Kolmakov}}, \bibinfo
  {author} {\bibfnamefont {L.~P.}\ \bibnamefont {Mezhov-Deglin}},\ and\
  \bibinfo {author} {\bibfnamefont {P.~V.~E.}\ \bibnamefont {McClintock}},\
  }\bibfield  {title} {\bibinfo {title} {Observation of an inverse energy
  cascade in developed acoustic turbulence in superfluid helium},\ }\href
  {https://doi.org/10.1103/PhysRevLett.101.065303} {\bibfield  {journal}
  {\bibinfo  {journal} {Phys. Rev. Lett.}\ }\textbf {\bibinfo {volume} {101}},\
  \bibinfo {pages} {065303} (\bibinfo {year} {2008})}\BibitemShut {NoStop}%
\bibitem [{\citenamefont {Solli}\ \emph {et~al.}(2007)\citenamefont {Solli},
  \citenamefont {Ropers}, \citenamefont {Koonath},\ and\ \citenamefont
  {Jalali}}]{Solli2007}%
  \BibitemOpen
  \bibfield  {author} {\bibinfo {author} {\bibfnamefont {D.~R.}\ \bibnamefont
  {Solli}}, \bibinfo {author} {\bibfnamefont {C.}~\bibnamefont {Ropers}},
  \bibinfo {author} {\bibfnamefont {P.}~\bibnamefont {Koonath}},\ and\ \bibinfo
  {author} {\bibfnamefont {B.}~\bibnamefont {Jalali}},\ }\bibfield  {title}
  {\bibinfo {title} {Optical rogue waves},\ }\href
  {https://doi.org/10.1038/nature06402} {\bibfield  {journal} {\bibinfo
  {journal} {Nature}\ }\textbf {\bibinfo {volume} {450}},\ \bibinfo {pages}
  {1054} (\bibinfo {year} {2007})}\BibitemShut {NoStop}%
\bibitem [{\citenamefont {Buccoliero}\ \emph {et~al.}(2011)\citenamefont
  {Buccoliero}, \citenamefont {Steffensen}, \citenamefont
  {Ebendorff-Heidepriem}, \citenamefont {Monro},\ and\ \citenamefont
  {Bang}}]{Buccoliero2011}%
  \BibitemOpen
  \bibfield  {author} {\bibinfo {author} {\bibfnamefont {D.}~\bibnamefont
  {Buccoliero}}, \bibinfo {author} {\bibfnamefont {H.}~\bibnamefont
  {Steffensen}}, \bibinfo {author} {\bibfnamefont {H.}~\bibnamefont
  {Ebendorff-Heidepriem}}, \bibinfo {author} {\bibfnamefont {T.~M.}\
  \bibnamefont {Monro}},\ and\ \bibinfo {author} {\bibfnamefont
  {O.}~\bibnamefont {Bang}},\ }\bibfield  {title} {\bibinfo {title}
  {Midinfrared optical rogue waves in soft glass photonic crystal fiber},\
  }\href {https://doi.org/10.1364/OE.19.017973} {\bibfield  {journal} {\bibinfo
   {journal} {Opt. Express}\ }\textbf {\bibinfo {volume} {19}},\ \bibinfo
  {pages} {17973} (\bibinfo {year} {2011})}\BibitemShut {NoStop}%
\bibitem [{\citenamefont {Onorato}\ \emph {et~al.}(2013)\citenamefont
  {Onorato}, \citenamefont {Residori}, \citenamefont {Bortolozzo},
  \citenamefont {Montina},\ and\ \citenamefont {Arecchi}}]{Onorato2013}%
  \BibitemOpen
  \bibfield  {author} {\bibinfo {author} {\bibfnamefont {M.}~\bibnamefont
  {Onorato}}, \bibinfo {author} {\bibfnamefont {S.}~\bibnamefont {Residori}},
  \bibinfo {author} {\bibfnamefont {U.}~\bibnamefont {Bortolozzo}}, \bibinfo
  {author} {\bibfnamefont {A.}~\bibnamefont {Montina}},\ and\ \bibinfo {author}
  {\bibfnamefont {F.}~\bibnamefont {Arecchi}},\ }\bibfield  {title} {\bibinfo
  {title} {Rogue waves and their generating mechanisms in different physical
  contexts},\ }\href
  {https://doi.org/https://doi.org/10.1016/j.physrep.2013.03.001} {\bibfield
  {journal} {\bibinfo  {journal} {Physics Reports}\ }\textbf {\bibinfo {volume}
  {528}},\ \bibinfo {pages} {47} (\bibinfo {year} {2013})},\ \bibinfo {note}
  {rogue waves and their generating mechanisms in different physical
  contexts}\BibitemShut {NoStop}%
\bibitem [{\citenamefont {Hansen}\ \emph {et~al.}(2021)\citenamefont {Hansen},
  \citenamefont {Engelsholm}, \citenamefont {Petersen},\ and\ \citenamefont
  {Bang}}]{Hansen2021}%
  \BibitemOpen
  \bibfield  {author} {\bibinfo {author} {\bibfnamefont {R.~E.}\ \bibnamefont
  {Hansen}}, \bibinfo {author} {\bibfnamefont {R.~D.}\ \bibnamefont
  {Engelsholm}}, \bibinfo {author} {\bibfnamefont {C.~R.}\ \bibnamefont
  {Petersen}},\ and\ \bibinfo {author} {\bibfnamefont {O.}~\bibnamefont
  {Bang}},\ }\bibfield  {title} {\bibinfo {title} {Numerical observation of spm
  rogue waves in normal dispersion cascaded supercontinuum generation},\ }\href
  {https://doi.org/10.1364/JOSAB.428520} {\bibfield  {journal} {\bibinfo
  {journal} {J. Opt. Soc. Am. B}\ }\textbf {\bibinfo {volume} {38}},\ \bibinfo
  {pages} {2754} (\bibinfo {year} {2021})}\BibitemShut {NoStop}%
\bibitem [{\citenamefont {Pareto}(1896)}]{Pareto1896}%
  \BibitemOpen
  \bibfield  {author} {\bibinfo {author} {\bibfnamefont {V.}~\bibnamefont
  {Pareto}},\ }\href@noop {} {\emph {\bibinfo {title} {Cours d'{\'e}conomie
  politique profess{\'e} {\`a} l'Universit{\'e} de Lausanne.}}}\ (\bibinfo
  {publisher} {F. Rouge; Pichon},\ \bibinfo {address} {Lausanne; Paris},\
  \bibinfo {year} {1896})\BibitemShut {NoStop}%
\bibitem [{\citenamefont {Yakovenko}\ and\ \citenamefont
  {Rosser}(2009)}]{Yakovenko2009}%
  \BibitemOpen
  \bibfield  {author} {\bibinfo {author} {\bibfnamefont {V.~M.}\ \bibnamefont
  {Yakovenko}}\ and\ \bibinfo {author} {\bibfnamefont {J.~B.}\ \bibnamefont
  {Rosser}},\ }\bibfield  {title} {\bibinfo {title} {Colloquium: Statistical
  mechanics of money, wealth, and income},\ }\href
  {https://doi.org/10.1103/RevModPhys.81.1703} {\bibfield  {journal} {\bibinfo
  {journal} {Rev. Mod. Phys.}\ }\textbf {\bibinfo {volume} {81}},\ \bibinfo
  {pages} {1703} (\bibinfo {year} {2009})}\BibitemShut {NoStop}%
\bibitem [{\citenamefont {Shpilberg}(1977)}]{Shpilberg1977}%
  \BibitemOpen
  \bibfield  {author} {\bibinfo {author} {\bibfnamefont {D.~C.}\ \bibnamefont
  {Shpilberg}},\ }\bibfield  {title} {\bibinfo {title} {The probability
  distribution of fire loss amount},\ }\href
  {http://www.jstor.org/stable/251860} {\bibfield  {journal} {\bibinfo
  {journal} {The Journal of Risk and Insurance}\ }\textbf {\bibinfo {volume}
  {44}},\ \bibinfo {pages} {103} (\bibinfo {year} {1977})}\BibitemShut
  {NoStop}%
\bibitem [{\citenamefont {Rootzén}\ and\ \citenamefont
  {Tajvidi}(1995)}]{Rootzen1995}%
  \BibitemOpen
  \bibfield  {author} {\bibinfo {author} {\bibfnamefont {H.}~\bibnamefont
  {Rootzén}}\ and\ \bibinfo {author} {\bibfnamefont {N.}~\bibnamefont
  {Tajvidi}},\ }\bibfield  {title} {\bibinfo {title} {Extreme value statistics
  and wind storm losses: A case study},\ }\href@noop {} {\bibfield  {journal}
  {\bibinfo  {journal} {Scand. Actuarial J}\ }\textbf {\bibinfo {volume} {1}},\
  \bibinfo {pages} {70} (\bibinfo {year} {1995})}\BibitemShut {NoStop}%
\bibitem [{\citenamefont {de~Solla~Price}(1965)}]{Price1965}%
  \BibitemOpen
  \bibfield  {author} {\bibinfo {author} {\bibfnamefont {D.~J.}\ \bibnamefont
  {de~Solla~Price}},\ }\bibfield  {title} {\bibinfo {title} {Networks of
  scientific papers},\ }\href {https://doi.org/10.1126/science.149.3683.510}
  {\bibfield  {journal} {\bibinfo  {journal} {Science}\ }\textbf {\bibinfo
  {volume} {149}},\ \bibinfo {pages} {510} (\bibinfo {year}
  {1965})}\BibitemShut {NoStop}%
\bibitem [{\citenamefont {Redner}(1998)}]{Redner1998}%
  \BibitemOpen
  \bibfield  {author} {\bibinfo {author} {\bibfnamefont {S.}~\bibnamefont
  {Redner}},\ }\bibfield  {title} {\bibinfo {title} {How popular is your paper?
  an empirical study of the citation distribution},\ }\href
  {https://doi.org/10.1007/s100510050359} {\bibfield  {journal} {\bibinfo
  {journal} {Eur. Phys. J. B}\ }\textbf {\bibinfo {volume} {4}},\ \bibinfo
  {pages} {131} (\bibinfo {year} {1998})}\BibitemShut {NoStop}%
\bibitem [{\citenamefont {Golosovsky}(2017)}]{Golosovsky2017}%
  \BibitemOpen
  \bibfield  {author} {\bibinfo {author} {\bibfnamefont {M.}~\bibnamefont
  {Golosovsky}},\ }\bibfield  {title} {\bibinfo {title} {Power-law citation
  distributions are not scale-free},\ }\href
  {https://doi.org/10.1103/PhysRevE.96.032306} {\bibfield  {journal} {\bibinfo
  {journal} {Phys. Rev. E}\ }\textbf {\bibinfo {volume} {96}},\ \bibinfo
  {pages} {032306} (\bibinfo {year} {2017})}\BibitemShut {NoStop}%
\bibitem [{\citenamefont {Gutenberg}\ and\ \citenamefont
  {Richter}(1944)}]{Gutenberg1944}%
  \BibitemOpen
  \bibfield  {author} {\bibinfo {author} {\bibfnamefont {B.}~\bibnamefont
  {Gutenberg}}\ and\ \bibinfo {author} {\bibfnamefont {C.~F.}\ \bibnamefont
  {Richter}},\ }\bibfield  {title} {\bibinfo {title} {{Frequency of earthquakes
  in California*}},\ }\href@noop {} {\bibfield  {journal} {\bibinfo  {journal}
  {Bulletin of the Seismological Society of America}\ }\textbf {\bibinfo
  {volume} {34}},\ \bibinfo {pages} {185} (\bibinfo {year} {1944})},\ \Eprint
  {https://arxiv.org/abs/https://authors.library.caltech.edu/47734/1/185.full.pdf}
  {https://authors.library.caltech.edu/47734/1/185.full.pdf} \BibitemShut
  {NoStop}%
\bibitem [{\citenamefont {Christensen}\ \emph {et~al.}(2002)\citenamefont
  {Christensen}, \citenamefont {Danon}, \citenamefont {Scanlon},\ and\
  \citenamefont {Bak}}]{Christensen2002}%
  \BibitemOpen
  \bibfield  {author} {\bibinfo {author} {\bibfnamefont {K.}~\bibnamefont
  {Christensen}}, \bibinfo {author} {\bibfnamefont {L.}~\bibnamefont {Danon}},
  \bibinfo {author} {\bibfnamefont {T.}~\bibnamefont {Scanlon}},\ and\ \bibinfo
  {author} {\bibfnamefont {P.}~\bibnamefont {Bak}},\ }\bibfield  {title}
  {\bibinfo {title} {Unified scaling law for earthquakes},\ }\href
  {https://doi.org/10.1073/pnas.012581099} {\bibfield  {journal} {\bibinfo
  {journal} {Proceedings of the National Academy of Sciences}\ }\textbf
  {\bibinfo {volume} {99}},\ \bibinfo {pages} {2509} (\bibinfo {year}
  {2002})}\BibitemShut {NoStop}%
\bibitem [{\citenamefont {Newberry}\ and\ \citenamefont
  {Savage}(2019)}]{Newberry2019}%
  \BibitemOpen
  \bibfield  {author} {\bibinfo {author} {\bibfnamefont {M.~G.}\ \bibnamefont
  {Newberry}}\ and\ \bibinfo {author} {\bibfnamefont {V.~M.}\ \bibnamefont
  {Savage}},\ }\bibfield  {title} {\bibinfo {title} {Self-similar processes
  follow a power law in discrete logarithmic space},\ }\href
  {https://doi.org/10.1103/PhysRevLett.122.158303} {\bibfield  {journal}
  {\bibinfo  {journal} {Phys. Rev. Lett.}\ }\textbf {\bibinfo {volume} {122}},\
  \bibinfo {pages} {158303} (\bibinfo {year} {2019})}\BibitemShut {NoStop}%
\bibitem [{\citenamefont {Birkeland}\ and\ \citenamefont
  {Landry}(2002)}]{Birkeland2002}%
  \BibitemOpen
  \bibfield  {author} {\bibinfo {author} {\bibfnamefont {K.~W.}\ \bibnamefont
  {Birkeland}}\ and\ \bibinfo {author} {\bibfnamefont {C.~C.}\ \bibnamefont
  {Landry}},\ }\bibfield  {title} {\bibinfo {title} {Power-laws and snow
  avalanches},\ }\href {https://doi.org/10.1029/2001GL014623} {\bibfield
  {journal} {\bibinfo  {journal} {Geophysical Research Letters}\ }\textbf
  {\bibinfo {volume} {29}},\ \bibinfo {pages} {49} (\bibinfo {year}
  {2002})}\BibitemShut {NoStop}%
\bibitem [{\citenamefont {Lu}\ and\ \citenamefont {Hamilton}(1991)}]{Lu1991}%
  \BibitemOpen
  \bibfield  {author} {\bibinfo {author} {\bibfnamefont {E.~T.}\ \bibnamefont
  {Lu}}\ and\ \bibinfo {author} {\bibfnamefont {R.~J.}\ \bibnamefont
  {Hamilton}},\ }\bibfield  {title} {\bibinfo {title} {Avalanches and the
  distribution of solar flares},\ }\href@noop {} {\bibfield  {journal}
  {\bibinfo  {journal} {The astrophysical journal}\ }\textbf {\bibinfo {volume}
  {380}},\ \bibinfo {pages} {L89} (\bibinfo {year} {1991})}\BibitemShut
  {NoStop}%
\bibitem [{\citenamefont {Pastor-Satorras}\ \emph {et~al.}(2001)\citenamefont
  {Pastor-Satorras}, \citenamefont {V\'azquez},\ and\ \citenamefont
  {Vespignani}}]{Pastor2001}%
  \BibitemOpen
  \bibfield  {author} {\bibinfo {author} {\bibfnamefont {R.}~\bibnamefont
  {Pastor-Satorras}}, \bibinfo {author} {\bibfnamefont {A.}~\bibnamefont
  {V\'azquez}},\ and\ \bibinfo {author} {\bibfnamefont {A.}~\bibnamefont
  {Vespignani}},\ }\bibfield  {title} {\bibinfo {title} {Dynamical and
  correlation properties of the internet},\ }\href
  {https://doi.org/10.1103/PhysRevLett.87.258701} {\bibfield  {journal}
  {\bibinfo  {journal} {Phys. Rev. Lett.}\ }\textbf {\bibinfo {volume} {87}},\
  \bibinfo {pages} {258701} (\bibinfo {year} {2001})}\BibitemShut {NoStop}%
\bibitem [{\citenamefont {Stumpf}\ and\ \citenamefont
  {Ingram}(2005)}]{Stumpf2005}%
  \BibitemOpen
  \bibfield  {author} {\bibinfo {author} {\bibfnamefont {M.~P.~H.}\
  \bibnamefont {Stumpf}}\ and\ \bibinfo {author} {\bibfnamefont {P.~J.}\
  \bibnamefont {Ingram}},\ }\bibfield  {title} {\bibinfo {title} {Probability
  models for degree distributions of protein interaction networks},\ }\href
  {https://doi.org/10.1209/epl/i2004-10531-8} {\bibfield  {journal} {\bibinfo
  {journal} {Europhysics Letters ({EPL})}\ }\textbf {\bibinfo {volume} {71}},\
  \bibinfo {pages} {152} (\bibinfo {year} {2005})}\BibitemShut {NoStop}%
\bibitem [{\citenamefont {Newman}(2005)}]{Newman2005}%
  \BibitemOpen
  \bibfield  {author} {\bibinfo {author} {\bibfnamefont {M.}~\bibnamefont
  {Newman}},\ }\bibfield  {title} {\bibinfo {title} {Power laws, {P}areto
  distributions and {Z}ipf's law},\ }\href
  {https://doi.org/10.1080/00107510500052444} {\bibfield  {journal} {\bibinfo
  {journal} {Contemporary Physics}\ }\textbf {\bibinfo {volume} {46}},\
  \bibinfo {pages} {323} (\bibinfo {year} {2005})}\BibitemShut {NoStop}%
\bibitem [{\citenamefont {Zipf}(1949)}]{Zipf1949}%
  \BibitemOpen
  \bibfield  {author} {\bibinfo {author} {\bibfnamefont {G.~K.}\ \bibnamefont
  {Zipf}},\ }\href@noop {} {\emph {\bibinfo {title} {Human behavior and the
  principle of least effort.}}},\ Human behavior and the principle of least
  effort.\ (\bibinfo  {publisher} {Addison-Wesley Press},\ \bibinfo {address}
  {Oxford, England},\ \bibinfo {year} {1949})\ pp.\ \bibinfo {pages} {xi,
  573--xi, 573}\BibitemShut {NoStop}%
\bibitem [{\citenamefont {Mandelbrot}(1960)}]{Mandelbrot1960}%
  \BibitemOpen
  \bibfield  {author} {\bibinfo {author} {\bibfnamefont {B.}~\bibnamefont
  {Mandelbrot}},\ }\bibfield  {title} {\bibinfo {title} {The {P}areto--{L}évy
  law and the distribution of income},\ }\href
  {http://www.jstor.org/stable/2525289} {\bibfield  {journal} {\bibinfo
  {journal} {International Economic Review}\ }\textbf {\bibinfo {volume} {1}},\
  \bibinfo {pages} {79} (\bibinfo {year} {1960})}\BibitemShut {NoStop}%
\bibitem [{\citenamefont {Mandelbrot}(1963)}]{Mandelbrot1963}%
  \BibitemOpen
  \bibfield  {author} {\bibinfo {author} {\bibfnamefont {B.}~\bibnamefont
  {Mandelbrot}},\ }\bibfield  {title} {\bibinfo {title} {New methods in
  statistical economics},\ }\href@noop {} {\bibfield  {journal} {\bibinfo
  {journal} {The Journal of Political Economy}\ }\textbf {\bibinfo {volume}
  {71}} (\bibinfo {year} {1963})}\BibitemShut {NoStop}%
\bibitem [{\citenamefont {Embrechts}\ \emph {et~al.}(1999)\citenamefont
  {Embrechts}, \citenamefont {Resnick},\ and\ \citenamefont
  {Samorodnitsky}}]{Embrechts1999}%
  \BibitemOpen
  \bibfield  {author} {\bibinfo {author} {\bibfnamefont {P.}~\bibnamefont
  {Embrechts}}, \bibinfo {author} {\bibfnamefont {S.~I.}\ \bibnamefont
  {Resnick}},\ and\ \bibinfo {author} {\bibfnamefont {G.}~\bibnamefont
  {Samorodnitsky}},\ }\bibfield  {title} {\bibinfo {title} {Extreme value
  theory as a risk management tool},\ }\href
  {https://doi.org/10.1080/10920277.1999.10595797} {\bibfield  {journal}
  {\bibinfo  {journal} {North American Actuarial Journal}\ }\textbf {\bibinfo
  {volume} {3}},\ \bibinfo {pages} {30} (\bibinfo {year} {1999})}\BibitemShut
  {NoStop}%
\bibitem [{\citenamefont {Rachev}(2003)}]{Rachev2003}%
  \BibitemOpen
  \bibinfo {editor} {\bibfnamefont {S.}~\bibnamefont {Rachev}},\ ed.,\ \href
  {https://ideas.repec.org/b/eee/monogr/9780444508966.html} {\emph {\bibinfo
  {title} {{Handbook of Heavy Tailed Distributions in Finance}}}},\ \bibinfo
  {series} {Elsevier Monographs}\ No.\ \bibinfo {number} {9780444508966}\
  (\bibinfo  {publisher} {Elsevier},\ \bibinfo {year} {2003})\BibitemShut
  {NoStop}%
\bibitem [{\citenamefont {Akhmediev}\ and\ \citenamefont
  {Pelinovsky}(2010)}]{Akhmediev2010}%
  \BibitemOpen
  \bibfield  {author} {\bibinfo {author} {\bibfnamefont {N.}~\bibnamefont
  {Akhmediev}}\ and\ \bibinfo {author} {\bibfnamefont {E.}~\bibnamefont
  {Pelinovsky}},\ }\bibfield  {title} {\bibinfo {title} {{E}ditorial --
  {I}ntroductory remarks on ``{D}iscussion {\&} {D}ebate: {R}ogue waves --
  {T}owards a {U}nifying {C}oncept?''},\ }\href
  {https://doi.org/10.1140/epjst/e2010-01233-0} {\bibfield  {journal} {\bibinfo
   {journal} {The European Physical Journal Special Topics}\ }\textbf {\bibinfo
  {volume} {185}},\ \bibinfo {pages} {1} (\bibinfo {year} {2010})}\BibitemShut
  {NoStop}%
\bibitem [{\citenamefont {Hill}(1975)}]{Hill1975}%
  \BibitemOpen
  \bibfield  {author} {\bibinfo {author} {\bibfnamefont {B.~M.}\ \bibnamefont
  {Hill}},\ }\bibfield  {title} {\bibinfo {title} {A simple general approach to
  inference about the tail of a distribution},\ }\href
  {https://doi.org/10.1214/aos/1176343247} {\bibfield  {journal} {\bibinfo
  {journal} {Ann. Statist.}\ }\textbf {\bibinfo {volume} {3}},\ \bibinfo
  {pages} {1163} (\bibinfo {year} {1975})}\BibitemShut {NoStop}%
\bibitem [{\citenamefont {III}(1975)}]{Pickands1975}%
  \BibitemOpen
  \bibfield  {author} {\bibinfo {author} {\bibfnamefont {J.~P.}\ \bibnamefont
  {III}},\ }\bibfield  {title} {\bibinfo {title} {{Statistical Inference Using
  Extreme Order Statistics}},\ }\href {https://doi.org/10.1214/aos/1176343003}
  {\bibfield  {journal} {\bibinfo  {journal} {The Annals of Statistics}\
  }\textbf {\bibinfo {volume} {3}},\ \bibinfo {pages} {119 } (\bibinfo {year}
  {1975})}\BibitemShut {NoStop}%
\bibitem [{\citenamefont {Kratz}\ and\ \citenamefont
  {Resnick}(1996)}]{Kratz1996}%
  \BibitemOpen
  \bibfield  {author} {\bibinfo {author} {\bibfnamefont {M.}~\bibnamefont
  {Kratz}}\ and\ \bibinfo {author} {\bibfnamefont {S.~I.}\ \bibnamefont
  {Resnick}},\ }\bibfield  {title} {\bibinfo {title} {The qq-estimator and
  heavy tails},\ }\href {https://doi.org/10.1080/15326349608807407} {\bibfield
  {journal} {\bibinfo  {journal} {Communications in Statistics. Stochastic
  Models}\ }\textbf {\bibinfo {volume} {12}},\ \bibinfo {pages} {699} (\bibinfo
  {year} {1996})}\BibitemShut {NoStop}%
\bibitem [{\citenamefont {Feuerverger}\ and\ \citenamefont
  {Hall}(1999)}]{Feuerverger1999}%
  \BibitemOpen
  \bibfield  {author} {\bibinfo {author} {\bibfnamefont {A.}~\bibnamefont
  {Feuerverger}}\ and\ \bibinfo {author} {\bibfnamefont {P.}~\bibnamefont
  {Hall}},\ }\bibfield  {title} {\bibinfo {title} {{Estimating a tail exponent
  by modelling departure from a {P}areto distribution}},\ }\href
  {https://doi.org/10.1214/aos/1018031215} {\bibfield  {journal} {\bibinfo
  {journal} {The Annals of Statistics}\ }\textbf {\bibinfo {volume} {27}},\
  \bibinfo {pages} {760 } (\bibinfo {year} {1999})}\BibitemShut {NoStop}%
\bibitem [{\citenamefont {Boyd}(2008)}]{Boyd2008}%
  \BibitemOpen
  \bibfield  {author} {\bibinfo {author} {\bibfnamefont {R.~W.}\ \bibnamefont
  {Boyd}},\ }\href@noop {} {\emph {\bibinfo {title} {Nonlinear Optics, Third
  Edition}}},\ \bibinfo {edition} {3rd}\ ed.\ (\bibinfo  {publisher} {Academic
  Press, Inc.},\ \bibinfo {address} {USA},\ \bibinfo {year} {2008})\BibitemShut
  {NoStop}%
\bibitem [{\citenamefont {Grynberg}\ \emph {et~al.}(2010)\citenamefont
  {Grynberg}, \citenamefont {Aspect},\ and\ \citenamefont
  {Fabre}}]{Grynberg2010}%
  \BibitemOpen
  \bibfield  {author} {\bibinfo {author} {\bibfnamefont {G.}~\bibnamefont
  {Grynberg}}, \bibinfo {author} {\bibfnamefont {A.}~\bibnamefont {Aspect}},\
  and\ \bibinfo {author} {\bibfnamefont {C.}~\bibnamefont {Fabre}},\ }\href
  {https://doi.org/10.1017/CBO9780511778261} {\emph {\bibinfo {title}
  {Introduction to Quantum Optics: From the Semi-classical Approach to
  Quantized Light}}}\ (\bibinfo  {publisher} {Cambridge University Press},\
  \bibinfo {address} {Cambridge},\ \bibinfo {year} {2010})\BibitemShut
  {NoStop}%
\bibitem [{\citenamefont {Barthelemy}\ \emph {et~al.}(2008)\citenamefont
  {Barthelemy}, \citenamefont {Bertolotti},\ and\ \citenamefont
  {Wiersma}}]{Barthelemy2008}%
  \BibitemOpen
  \bibfield  {author} {\bibinfo {author} {\bibfnamefont {P.}~\bibnamefont
  {Barthelemy}}, \bibinfo {author} {\bibfnamefont {J.}~\bibnamefont
  {Bertolotti}},\ and\ \bibinfo {author} {\bibfnamefont {D.~S.}\ \bibnamefont
  {Wiersma}},\ }\bibfield  {title} {\bibinfo {title} {A {L}{\'e}vy flight for
  light},\ }\href {https://doi.org/10.1038/nature06948} {\bibfield  {journal}
  {\bibinfo  {journal} {Nature}\ }\textbf {\bibinfo {volume} {453}},\ \bibinfo
  {pages} {495} (\bibinfo {year} {2008})}\BibitemShut {NoStop}%
\bibitem [{\citenamefont {Mercadier}\ \emph {et~al.}(2009)\citenamefont
  {Mercadier}, \citenamefont {Guerin}, \citenamefont {Chevrollier},\ and\
  \citenamefont {Kaiser}}]{Mercadier2009}%
  \BibitemOpen
  \bibfield  {author} {\bibinfo {author} {\bibfnamefont {N.}~\bibnamefont
  {Mercadier}}, \bibinfo {author} {\bibfnamefont {W.}~\bibnamefont {Guerin}},
  \bibinfo {author} {\bibfnamefont {M.}~\bibnamefont {Chevrollier}},\ and\
  \bibinfo {author} {\bibfnamefont {R.}~\bibnamefont {Kaiser}},\ }\bibfield
  {title} {\bibinfo {title} {L{\'e}vy flights of photons in hot atomic
  vapours},\ }\href {https://doi.org/10.1038/nphys1286} {\bibfield  {journal}
  {\bibinfo  {journal} {Nature Physics}\ }\textbf {\bibinfo {volume} {5}},\
  \bibinfo {pages} {602} (\bibinfo {year} {2009})}\BibitemShut {NoStop}%
\bibitem [{\citenamefont {Solli}\ \emph {et~al.}(2008)\citenamefont {Solli},
  \citenamefont {Ropers},\ and\ \citenamefont {Jalali}}]{Solli2008}%
  \BibitemOpen
  \bibfield  {author} {\bibinfo {author} {\bibfnamefont {D.~R.}\ \bibnamefont
  {Solli}}, \bibinfo {author} {\bibfnamefont {C.}~\bibnamefont {Ropers}},\ and\
  \bibinfo {author} {\bibfnamefont {B.}~\bibnamefont {Jalali}},\ }\bibfield
  {title} {\bibinfo {title} {Active control of rogue waves for stimulated
  supercontinuum generation},\ }\href
  {https://doi.org/10.1103/PhysRevLett.101.233902} {\bibfield  {journal}
  {\bibinfo  {journal} {Phys. Rev. Lett.}\ }\textbf {\bibinfo {volume} {101}},\
  \bibinfo {pages} {233902} (\bibinfo {year} {2008})}\BibitemShut {NoStop}%
\bibitem [{\citenamefont {Wetzel}\ \emph
  {et~al.}(2012{\natexlab{a}})\citenamefont {Wetzel}, \citenamefont {Blow},
  \citenamefont {Turitsyn}, \citenamefont {Millot}, \citenamefont {Larger},\
  and\ \citenamefont {Dudley}}]{Wetzel2012b}%
  \BibitemOpen
  \bibfield  {author} {\bibinfo {author} {\bibfnamefont {B.}~\bibnamefont
  {Wetzel}}, \bibinfo {author} {\bibfnamefont {K.~J.}\ \bibnamefont {Blow}},
  \bibinfo {author} {\bibfnamefont {S.~K.}\ \bibnamefont {Turitsyn}}, \bibinfo
  {author} {\bibfnamefont {G.}~\bibnamefont {Millot}}, \bibinfo {author}
  {\bibfnamefont {L.}~\bibnamefont {Larger}},\ and\ \bibinfo {author}
  {\bibfnamefont {J.~M.}\ \bibnamefont {Dudley}},\ }\bibfield  {title}
  {\bibinfo {title} {Random walks and random numbers from supercontinuum
  generation},\ }\href {https://doi.org/10.1364/OE.20.011143} {\bibfield
  {journal} {\bibinfo  {journal} {Opt. Express}\ }\textbf {\bibinfo {volume}
  {20}},\ \bibinfo {pages} {11143} (\bibinfo {year}
  {2012}{\natexlab{a}})}\BibitemShut {NoStop}%
\bibitem [{\citenamefont {Sørensen}\ \emph {et~al.}(2012)\citenamefont
  {Sørensen}, \citenamefont {Bang}, \citenamefont {Wetzel},\ and\
  \citenamefont {Dudley}}]{Sorensen2012}%
  \BibitemOpen
  \bibfield  {author} {\bibinfo {author} {\bibfnamefont {S.~T.}\ \bibnamefont
  {Sørensen}}, \bibinfo {author} {\bibfnamefont {O.}~\bibnamefont {Bang}},
  \bibinfo {author} {\bibfnamefont {B.}~\bibnamefont {Wetzel}},\ and\ \bibinfo
  {author} {\bibfnamefont {J.~M.}\ \bibnamefont {Dudley}},\ }\bibfield  {title}
  {\bibinfo {title} {Describing supercontinuum noise and rogue wave statistics
  using higher-order moments},\ }\href
  {https://doi.org/https://doi.org/10.1016/j.optcom.2012.01.030} {\bibfield
  {journal} {\bibinfo  {journal} {Optics Communications}\ }\textbf {\bibinfo
  {volume} {285}},\ \bibinfo {pages} {2451} (\bibinfo {year}
  {2012})}\BibitemShut {NoStop}%
\bibitem [{\citenamefont {Wetzel}\ \emph
  {et~al.}(2012{\natexlab{b}})\citenamefont {Wetzel}, \citenamefont {Stefani},
  \citenamefont {Larger}, \citenamefont {Lacourt}, \citenamefont {Merolla},
  \citenamefont {Sylvestre}, \citenamefont {Kudlinski}, \citenamefont {Mussot},
  \citenamefont {Genty}, \citenamefont {Dias},\ and\ \citenamefont
  {Dudley}}]{Wetzel2012}%
  \BibitemOpen
  \bibfield  {author} {\bibinfo {author} {\bibfnamefont {B.}~\bibnamefont
  {Wetzel}}, \bibinfo {author} {\bibfnamefont {A.}~\bibnamefont {Stefani}},
  \bibinfo {author} {\bibfnamefont {L.}~\bibnamefont {Larger}}, \bibinfo
  {author} {\bibfnamefont {P.~A.}\ \bibnamefont {Lacourt}}, \bibinfo {author}
  {\bibfnamefont {J.~M.}\ \bibnamefont {Merolla}}, \bibinfo {author}
  {\bibfnamefont {T.}~\bibnamefont {Sylvestre}}, \bibinfo {author}
  {\bibfnamefont {A.}~\bibnamefont {Kudlinski}}, \bibinfo {author}
  {\bibfnamefont {A.}~\bibnamefont {Mussot}}, \bibinfo {author} {\bibfnamefont
  {G.}~\bibnamefont {Genty}}, \bibinfo {author} {\bibfnamefont
  {F.}~\bibnamefont {Dias}},\ and\ \bibinfo {author} {\bibfnamefont {J.~M.}\
  \bibnamefont {Dudley}},\ }\bibfield  {title} {\bibinfo {title} {Real-time
  full bandwidth measurement of spectral noise in supercontinuum generation},\
  }\href {https://doi.org/10.1038/srep00882} {\bibfield  {journal} {\bibinfo
  {journal} {Scientific Reports}\ }\textbf {\bibinfo {volume} {2}},\ \bibinfo
  {pages} {882} (\bibinfo {year} {2012}{\natexlab{b}})}\BibitemShut {NoStop}%
\bibitem [{\citenamefont {Manceau}\ \emph {et~al.}(2019)\citenamefont
  {Manceau}, \citenamefont {Spasibko}, \citenamefont {Leuchs}, \citenamefont
  {Filip},\ and\ \citenamefont {Chekhova}}]{Manceau2019}%
  \BibitemOpen
  \bibfield  {author} {\bibinfo {author} {\bibfnamefont {M.}~\bibnamefont
  {Manceau}}, \bibinfo {author} {\bibfnamefont {K.~Y.}\ \bibnamefont
  {Spasibko}}, \bibinfo {author} {\bibfnamefont {G.}~\bibnamefont {Leuchs}},
  \bibinfo {author} {\bibfnamefont {R.}~\bibnamefont {Filip}},\ and\ \bibinfo
  {author} {\bibfnamefont {M.~V.}\ \bibnamefont {Chekhova}},\ }\bibfield
  {title} {\bibinfo {title} {Indefinite-mean {P}areto photon distribution from
  amplified quantum noise},\ }\href
  {https://doi.org/10.1103/PhysRevLett.123.123606} {\bibfield  {journal}
  {\bibinfo  {journal} {Phys. Rev. Lett.}\ }\textbf {\bibinfo {volume} {123}},\
  \bibinfo {pages} {123606} (\bibinfo {year} {2019})}\BibitemShut {NoStop}%
\bibitem [{\citenamefont {Quimby}(2006)}]{Quimby2006}%
  \BibitemOpen
  \bibfield  {author} {\bibinfo {author} {\bibfnamefont {R.}~\bibnamefont
  {Quimby}},\ }\href {https://books.google.cz/books?id=82f-gIvtC7wC} {\emph
  {\bibinfo {title} {{P}hotonics and {L}asers: {A}n {I}ntroduction}}}\
  (\bibinfo  {publisher} {John Wiley \& Sons},\ \bibinfo {year}
  {2006})\BibitemShut {NoStop}%
\bibitem [{\citenamefont {Agrawal}(2012)}]{Agrawal2012}%
  \BibitemOpen
  \bibfield  {author} {\bibinfo {author} {\bibfnamefont {G.}~\bibnamefont
  {Agrawal}},\ }\href {https://books.google.cz/books?id=xNvw-GDVn84C} {\emph
  {\bibinfo {title} {{N}onlinear {F}iber {O}ptics}}},\ Optics and Photonics\
  (\bibinfo  {publisher} {{E}lsevier {S}cience},\ \bibinfo {year}
  {2012})\BibitemShut {NoStop}%
\bibitem [{\citenamefont {Fedotov}\ \emph {et~al.}(2003)\citenamefont
  {Fedotov}, \citenamefont {Bugar}, \citenamefont {Sidorov-Biryukov},
  \citenamefont {Serebryannikov}, \citenamefont {Chorvat~Jr.}, \citenamefont
  {Scalora}, \citenamefont {Chorvat},\ and\ \citenamefont
  {Zheltikov}}]{Fedotov2003}%
  \BibitemOpen
  \bibfield  {author} {\bibinfo {author} {\bibfnamefont {A.~B.}\ \bibnamefont
  {Fedotov}}, \bibinfo {author} {\bibfnamefont {I.}~\bibnamefont {Bugar}},
  \bibinfo {author} {\bibfnamefont {D.~A.}\ \bibnamefont {Sidorov-Biryukov}},
  \bibinfo {author} {\bibfnamefont {E.~E.}\ \bibnamefont {Serebryannikov}},
  \bibinfo {author} {\bibfnamefont {D.}~\bibnamefont {Chorvat~Jr.}}, \bibinfo
  {author} {\bibfnamefont {M.}~\bibnamefont {Scalora}}, \bibinfo {author}
  {\bibfnamefont {D.}~\bibnamefont {Chorvat}},\ and\ \bibinfo {author}
  {\bibfnamefont {A.~M.}\ \bibnamefont {Zheltikov}},\ }\bibfield  {title}
  {\bibinfo {title} {Pump-depleting four-wave mixing in
  supercontinuum-generating microstructure fibers},\ }\href
  {https://doi.org/10.1007/s00340-003-1206-7} {\bibfield  {journal} {\bibinfo
  {journal} {Applied Physics B}\ }\textbf {\bibinfo {volume} {77}},\ \bibinfo
  {pages} {313} (\bibinfo {year} {2003})}\BibitemShut {NoStop}%
\bibitem [{\citenamefont {Vanholsbeeck}\ \emph {et~al.}(2005)\citenamefont
  {Vanholsbeeck}, \citenamefont {Martin-Lopez}, \citenamefont
  {Gonz\'{a}lez-Herr\'{a}ez},\ and\ \citenamefont {Coen}}]{Vanholsbeeck2005}%
  \BibitemOpen
  \bibfield  {author} {\bibinfo {author} {\bibfnamefont {F.}~\bibnamefont
  {Vanholsbeeck}}, \bibinfo {author} {\bibfnamefont {S.}~\bibnamefont
  {Martin-Lopez}}, \bibinfo {author} {\bibfnamefont {M.}~\bibnamefont
  {Gonz\'{a}lez-Herr\'{a}ez}},\ and\ \bibinfo {author} {\bibfnamefont
  {S.}~\bibnamefont {Coen}},\ }\bibfield  {title} {\bibinfo {title} {The role
  of pump incoherence in continuous-wave supercontinuum generation},\ }\href
  {https://doi.org/10.1364/OPEX.13.006615} {\bibfield  {journal} {\bibinfo
  {journal} {Opt. Express}\ }\textbf {\bibinfo {volume} {13}},\ \bibinfo
  {pages} {6615} (\bibinfo {year} {2005})}\BibitemShut {NoStop}%
\bibitem [{\citenamefont {R{\'{a}}cz}\ \emph {et~al.}(2021)\citenamefont
  {R{\'{a}}cz}, \citenamefont {Ruppert},\ and\ \citenamefont
  {Filip}}]{Racz2021}%
  \BibitemOpen
  \bibfield  {author} {\bibinfo {author} {\bibfnamefont {{\'{E}}.}~\bibnamefont
  {R{\'{a}}cz}}, \bibinfo {author} {\bibfnamefont {L.}~\bibnamefont
  {Ruppert}},\ and\ \bibinfo {author} {\bibfnamefont {R.}~\bibnamefont
  {Filip}},\ }\bibfield  {title} {\bibinfo {title} {Estimation of heavy tails
  in optical non-linear processes},\ }\href
  {https://doi.org/10.1088/1367-2630/abe442} {\bibfield  {journal} {\bibinfo
  {journal} {New Journal of Physics}\ }\textbf {\bibinfo {volume} {23}},\
  \bibinfo {pages} {043013} (\bibinfo {year} {2021})}\BibitemShut {NoStop}%
\bibitem [{\citenamefont {Spearman}(1904)}]{Spearman1904}%
  \BibitemOpen
  \bibfield  {author} {\bibinfo {author} {\bibfnamefont {C.}~\bibnamefont
  {Spearman}},\ }\bibfield  {title} {\bibinfo {title} {The proof and
  measurement of association between two things},\ }\href
  {http://www.jstor.org/stable/1412159} {\bibfield  {journal} {\bibinfo
  {journal} {The American Journal of Psychology}\ }\textbf {\bibinfo {volume}
  {15}},\ \bibinfo {pages} {72} (\bibinfo {year} {1904})}\BibitemShut {NoStop}%
\bibitem [{\citenamefont {Hollander}\ \emph {et~al.}(2013)\citenamefont
  {Hollander}, \citenamefont {Wolfe},\ and\ \citenamefont
  {Chicken}}]{Hollander2013}%
  \BibitemOpen
  \bibfield  {author} {\bibinfo {author} {\bibfnamefont {M.}~\bibnamefont
  {Hollander}}, \bibinfo {author} {\bibfnamefont {D.~A.}\ \bibnamefont
  {Wolfe}},\ and\ \bibinfo {author} {\bibfnamefont {E.}~\bibnamefont
  {Chicken}},\ }\href@noop {} {\emph {\bibinfo {title} {Nonparametric
  statistical methods}}},\ Vol.\ \bibinfo {volume} {751}\ (\bibinfo
  {publisher} {John Wiley \& Sons},\ \bibinfo {year} {2013})\BibitemShut
  {NoStop}%
\bibitem [{Note1()}]{Note1}%
  \BibitemOpen
  \bibinfo {note} {Note that the definition would be the same if one were to
  use the distribution function \(F(x) \equiv 1-\protect \overline F(x)\) in
  the formula above.}\BibitemShut {Stop}%
\bibitem [{\citenamefont {Drees}\ and\ \citenamefont
  {Kaufmann}(1998)}]{Drees1998}%
  \BibitemOpen
  \bibfield  {author} {\bibinfo {author} {\bibfnamefont {H.}~\bibnamefont
  {Drees}}\ and\ \bibinfo {author} {\bibfnamefont {E.}~\bibnamefont
  {Kaufmann}},\ }\bibfield  {title} {\bibinfo {title} {Selecting the optimal
  sample fraction in univariate extreme value estimation},\ }\href
  {https://doi.org/https://doi.org/10.1016/S0304-4149(98)00017-9} {\bibfield
  {journal} {\bibinfo  {journal} {Stochastic Processes and their Applications}\
  }\textbf {\bibinfo {volume} {75}},\ \bibinfo {pages} {149 } (\bibinfo {year}
  {1998})}\BibitemShut {NoStop}%
\bibitem [{\citenamefont {Guillou}\ and\ \citenamefont
  {Hall}(2001)}]{Guillou2001}%
  \BibitemOpen
  \bibfield  {author} {\bibinfo {author} {\bibfnamefont {A.}~\bibnamefont
  {Guillou}}\ and\ \bibinfo {author} {\bibfnamefont {P.}~\bibnamefont {Hall}},\
  }\bibfield  {title} {\bibinfo {title} {A diagnostic for selecting the
  threshold in extreme value analysis},\ }\href
  {https://doi.org/10.1111/1467-9868.00286} {\bibfield  {journal} {\bibinfo
  {journal} {Journal of the Royal Statistical Society: Series B (Statistical
  Methodology)}\ }\textbf {\bibinfo {volume} {63}},\ \bibinfo {pages} {293}
  (\bibinfo {year} {2001})}\BibitemShut {NoStop}%
\bibitem [{\citenamefont {Danielsson}\ \emph {et~al.}(2001)\citenamefont
  {Danielsson}, \citenamefont {de~Haan}, \citenamefont {Peng},\ and\
  \citenamefont {de~Vries}}]{Danielsson2001}%
  \BibitemOpen
  \bibfield  {author} {\bibinfo {author} {\bibfnamefont {J.}~\bibnamefont
  {Danielsson}}, \bibinfo {author} {\bibfnamefont {L.}~\bibnamefont {de~Haan}},
  \bibinfo {author} {\bibfnamefont {L.}~\bibnamefont {Peng}},\ and\ \bibinfo
  {author} {\bibfnamefont {C.}~\bibnamefont {de~Vries}},\ }\bibfield  {title}
  {\bibinfo {title} {Using a bootstrap method to choose the sample fraction in
  tail index estimation},\ }\href {https://doi.org/10.1006/jmva.2000.1903}
  {\bibfield  {journal} {\bibinfo  {journal} {J. Multivar. Anal.}\ }\textbf
  {\bibinfo {volume} {76}},\ \bibinfo {pages} {226–248} (\bibinfo {year}
  {2001})}\BibitemShut {NoStop}%
\bibitem [{\citenamefont {Caeiro}\ and\ \citenamefont
  {Gomes}(2015)}]{Caeiro2015}%
  \BibitemOpen
  \bibfield  {author} {\bibinfo {author} {\bibfnamefont {F.}~\bibnamefont
  {Caeiro}}\ and\ \bibinfo {author} {\bibfnamefont {M.}~\bibnamefont {Gomes}},\
  }\bibfield  {title} {\bibinfo {title} {Threshold selection in extreme value
  analysis: Methods and applications},\ }in\ \href
  {https://doi.org/10.1201/b19721-5} {\emph {\bibinfo {booktitle} {Extreme
  Value Modeling and Risk Analysis: Methods and Applications}}},\ \bibinfo
  {editor} {edited by\ \bibinfo {editor} {\bibfnamefont {D.}~\bibnamefont
  {Dey}}\ and\ \bibinfo {editor} {\bibfnamefont {J.}~\bibnamefont {Yan}}}\
  (\bibinfo  {publisher} {Taylor \& Francis},\ \bibinfo {address} {New York},\
  \bibinfo {year} {2015})\ pp.\ \bibinfo {pages} {69--86}\BibitemShut {NoStop}%
\bibitem [{\citenamefont {Neves}\ \emph {et~al.}(2015)\citenamefont {Neves},
  \citenamefont {Gomes}, \citenamefont {Figueiredo},\ and\ \citenamefont
  {Gomes}}]{Neves2015}%
  \BibitemOpen
  \bibfield  {author} {\bibinfo {author} {\bibfnamefont {M.~M.}\ \bibnamefont
  {Neves}}, \bibinfo {author} {\bibfnamefont {M.~I.}\ \bibnamefont {Gomes}},
  \bibinfo {author} {\bibfnamefont {F.}~\bibnamefont {Figueiredo}},\ and\
  \bibinfo {author} {\bibfnamefont {D.~P.}\ \bibnamefont {Gomes}},\ }\bibfield
  {title} {\bibinfo {title} {Modeling extreme events: Sample fraction adaptive
  choice in parameter estimation},\ }\href
  {https://doi.org/10.1080/15598608.2014.890984} {\bibfield  {journal}
  {\bibinfo  {journal} {Journal of Statistical Theory and Practice}\ }\textbf
  {\bibinfo {volume} {9}},\ \bibinfo {pages} {184} (\bibinfo {year}
  {2015})}\BibitemShut {NoStop}%
\bibitem [{Note2()}]{Note2}%
  \BibitemOpen
  \bibinfo {note} {Due to the \(\protect \qopname \relax o{sinh}^2(\cdot )\)
  transformation, this is essentially a multiplicative noise.}\BibitemShut
  {Stop}%
\bibitem [{\citenamefont {Drees}\ \emph {et~al.}(2000)\citenamefont {Drees},
  \citenamefont {de~Haan},\ and\ \citenamefont {Resnick}}]{Drees_HillPlot}%
  \BibitemOpen
  \bibfield  {author} {\bibinfo {author} {\bibfnamefont {H.}~\bibnamefont
  {Drees}}, \bibinfo {author} {\bibfnamefont {L.}~\bibnamefont {de~Haan}},\
  and\ \bibinfo {author} {\bibfnamefont {S.}~\bibnamefont {Resnick}},\
  }\bibfield  {title} {\bibinfo {title} {How to make a hill plot},\ }\href
  {http://www.jstor.org/stable/2673989} {\bibfield  {journal} {\bibinfo
  {journal} {The Annals of Statistics}\ }\textbf {\bibinfo {volume} {28}},\
  \bibinfo {pages} {254} (\bibinfo {year} {2000})}\BibitemShut {NoStop}%
\bibitem [{\citenamefont {Fl\'{o}rez}\ \emph {et~al.}(2020)\citenamefont
  {Fl\'{o}rez}, \citenamefont {Lundeen},\ and\ \citenamefont
  {Chekhova}}]{Florez2020}%
  \BibitemOpen
  \bibfield  {author} {\bibinfo {author} {\bibfnamefont {J.}~\bibnamefont
  {Fl\'{o}rez}}, \bibinfo {author} {\bibfnamefont {J.~S.}\ \bibnamefont
  {Lundeen}},\ and\ \bibinfo {author} {\bibfnamefont {M.~V.}\ \bibnamefont
  {Chekhova}},\ }\bibfield  {title} {\bibinfo {title} {Pump depletion in
  parametric down-conversion with low pump energies},\ }\href
  {https://doi.org/10.1364/OL.394925} {\bibfield  {journal} {\bibinfo
  {journal} {Opt. Lett.}\ }\textbf {\bibinfo {volume} {45}},\ \bibinfo {pages}
  {4264} (\bibinfo {year} {2020})}\BibitemShut {NoStop}%
\bibitem [{\citenamefont {Hand}\ and\ \citenamefont
  {Russell}(1990)}]{Hand1990}%
  \BibitemOpen
  \bibfield  {author} {\bibinfo {author} {\bibfnamefont {D.~P.}\ \bibnamefont
  {Hand}}\ and\ \bibinfo {author} {\bibfnamefont {P.~S.~J.}\ \bibnamefont
  {Russell}},\ }\bibfield  {title} {\bibinfo {title} {Photoinduced
  refractive-index changes in germanosilicate fibers},\ }\href
  {https://doi.org/10.1364/OL.15.000102} {\bibfield  {journal} {\bibinfo
  {journal} {Opt. Lett.}\ }\textbf {\bibinfo {volume} {15}},\ \bibinfo {pages}
  {102} (\bibinfo {year} {1990})}\BibitemShut {NoStop}%
\end{thebibliography}%

\end{document}